\documentclass[preprint]{elsarticle}
\usepackage{lineno}
\usepackage{graphicx}
\usepackage{bm}
\usepackage{epsf}
\usepackage{rotating}
\usepackage{epsfig,graphics,rotate,color}
\usepackage{amssymb}
\usepackage{amsmath}
\usepackage{amsfonts}
\usepackage{array,hhline,dcolumn}
\usepackage{multirow}
\providecommand{\U}[1]{\protect\rule{.1in}{.1in}}

\begin{document}


\title{A Prototype Detector for Directional Measurement of the Cosmogenic Neutron Flux}

\author{J. Lopez, K. Terao, J.M. Conrad, D. Dujmic, L. Winslow}
\address{Physics Dept., Massachusetts Institute of Technology,
Cambridge, MA 02139}

\begin{abstract}
  This paper describes a novel directional neutron
  detector. The low pressure time projection chamber uses a mix of
  helium and CF$_{4}$ gases.  The detector reconstructs the energy and
  angular distribution of fast neutron recoils.  This paper reports
  results of energy calibration using an $\alpha$ source and angular
  reconstruction studies using a collimated neutron source.  The best
  performance is obtained with a 12.5\% CF$_4$ -- 87.5\% He gas mixture.  At low
  energies the target for fast neutrons transitions is primarily
  helium, while at higher energies, the fluorine contributes as a target.  The reconstruction efficiency is both energy and target dependent.  For neutrons with energies less than 20~MeV, the reconstruction efficiency is $\sim$40\% for fluorine recoils and $\sim$60\% for helium recoils.

\end{abstract}
\maketitle

\section{Introduction}

Precise measurement of the energy and direction of
cosmogenic-muon-induced fast neutrons, with various levels of
underground shielding, will be valuable for future neutrino and dark
matter experiments \cite{JoeAnnualRev}.  In this paper, we
present a first-generation detector for this purpose.    The
detector design is based on that
for the directional dark matter search DMTPC \cite{DMTPC}.
We introduce modest modifications for directional neutron detection\cite{Roccaro:2009tg}.
Our first application will be in the Double Chooz
(DC) neutrino experiment \cite{DC}, 
and it is, therefore, called DCTPC.  

DCTPC is a low pressure time projection chamber (TPC) filled with two
gasses: (1) He, the fast neutron target, and (2) CF$_4$,
a scintillant and a quencher for the electron avalanche.
This detector is blind to x-rays and minimum ionizing particles (MIP), like
muons, because the density of primary ionization is too small to be detected.
The ionization electrons from a non-MIP particle track drift down to a stainless steel ground mesh. Between the ground mesh and the anode, an avalanche of electrons
occurs.  A charge-coupled device (CCD) camera installed in the detector images the
visible scintillation light from the avalanche.  The charge created in the avalanche is readout from the anode plate and ground mesh.  The track
direction and energy can be reconstructed using the CCD image and the charge
information. The reconstructed track from the recoiling nucleus can then be correlated 
to the incoming neutron direction and energy.
In this article we present on-surface measurements from a small 2.8~liter,
first-generation detector for an $\alpha$-source run and a
$^{252}$Cf neutron source run as well as an on-surface cosmogenic run.

\section{Motivation for DCTPC}

DCTPC provides a benchmark for the Double Chooz fast neutron simulation
at the specific sites of the two Double Chooz neutrino detectors \cite{DC}.  
The detectors sites 
have 114 and 300 meters water equivalent (m.w.e) shielding, respectively.
The final DCTPC design calls for
a large detector located at each site.  The small first-generation
detector 
described here will
be installed in the 300 m.w.e site for initial studies.  
The neutron
flux energy and angle measurements from the rock can be used to
tune the Monte Carlo and demonstrate a clear understanding of
this background in the Double Chooz neutrino oscillation analysis.
However, the results will provide wider benefits as
backgrounds from fast neutrons produced in rock are an
issue for all low background experiments. 

Low energy neutrons arise from  ($\alpha$, n) interactions due to natural radioactivity in the shielding rock.
At locations with low shielding, this
neutron source
is overwhelmed by neutron production from cosmic ray muons.
However, muons are attenuated with depth and  the ($\alpha$, n) background becomes
significant at 300~m.w.e. \cite{MeiAndHime}. 
DCTPC, with its very low energy threshold, can sample 
these ($\alpha$, n) events.
The
comparison of the 114 m.w.e. to 300 m.w.e. depths at Double Chooz will be useful in isolating
the ($\alpha$, n) from the muon-induced neutrons. 

The remaining two broad categories of neutron production are 
caused by cosmic ray muon interactions with rock:
stopped muon capture and deep inelastic
interactions.  The rates depend upon the depth of the final
experiment because low energy muons are attenuated with shielding.
DCTPC sensitivity will reach to $>$20 MeV in fast neutron
reconstruction, overlapping with the KARMEN2 (surface) and LVD (3200
m.w.e.) data sets.
As we show below, the maximum reconstructed recoil energy 
depends upon the size of the detector.

The FLUKA~\cite{fluka, fluka2} and GEANT4~\cite{G4, G42} simulations
are generally used to predict  $>10$ MeV fast neutrons that are produced by muons.  
DCTPC adds data points at two depths
which will complement the existing measurements.
Information on the energy
spectra of fast neutrons is scarce, coming only from KARMEN2
experiment on the surface~\cite{Karmen}, KamLAND at 2700
m.w.e.~\cite{KamLAND} and the LVD experiment at 3200
m.w.e.~\cite{Aglietta}.  
DCTPC
data will
fill the gap in depth between KARMEN2 and deeper experiments.
DCTPC
also provides an angular distribution of events.  In this paper, we 
show the quality of this angular reconstruction.  The unique directional
information can help disentangle different neutron sources.

\section{The First Generation Detector}

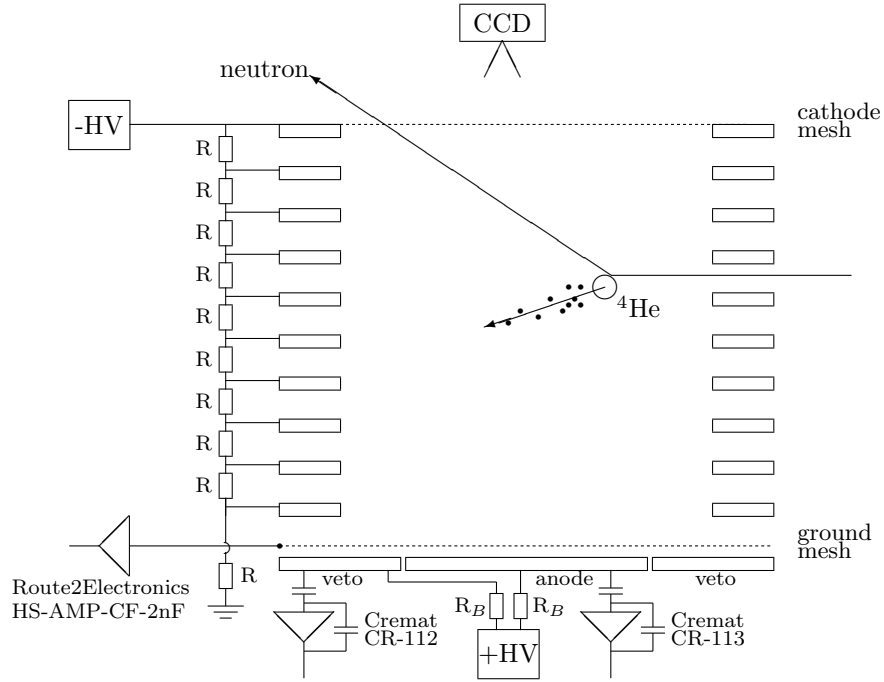
\begin{figure}[t]
\setlength{\unitlength}{0.8mm}
\begin{picture}(160,100)
\put(80,106){\framebox(14,6) {CCD}} 
\put(87,106){\line(-1,-2){3}}
\put(87,106){\line( 1,-2){3}}
\multiput(50,27)(0,7){10}{\framebox(10,2)}
\multiput(122,27)(0,7){10}{\framebox(10,2)}
\put(25,92.1){\line(1,0){35}}
\multiput(60,92.1)(1,0){68}{\line(1,0){0.5}}
\put(15,88){\framebox(10,8){-HV}}
\put(136,93){\small cathode}
\put(136,90){\small mesh}
\put(25,17){\line(0,1){10}} \put(25,27){\line(-1,-1){5}} \put(25,17){\line(-1,1){5}} \put(20,22){\line(-1,0){5}}
\put(25,22){\line(1,0){25}} \put(50,22){\circle*{1}} \multiput(50,22)(1,0){82}{\line(1,0){0.5}} 
\put(4,14){ \footnotesize Route2Electronics}
\put(4,10){ \footnotesize HS-AMP-CF-2nF}
\put(136,23){\small ground}
\put(136,20){\small mesh}
\multiput(40,30)(0,7){9}{\framebox(2,4)}
\multiput(41,30)(0,7){9}{\line(0,-1){3}}
\multiput(41,28.5)(0,7){9}{\line(1,0){9}} 
\put(41,92.1){\line(0,-1){2}}
\put(41,30){\line(0,-1){7.5}}
\put(41,21){\line(0,-1){2}}
\put(41.15,22){\oval(1,1)[r]}
\put(38,12){\line(1,0){6}}\put(39,11){\line(1,0){4}} \put(40,10){\line(1,0){2}}  
\put(40,15){\framebox(2,4)} \put(41,15){\line(0,-1){3}} \put(42,16){ \footnotesize R}
\multiput(36,31)(0,7){9}{\footnotesize R}
\put(50,18){\framebox(20,2)}  
\put(54,18){\line(0,-1){3}} 
\put(52,15){\line(1,0){4}} \put(52,14){\line(1,0){4}}
\put(54,14){\line(0,-1){3}} 
\put(49,11){\line(1,0){10}} \put(49,11){\line(1,-1){5}}  \put(59,11){\line(-1,-1){5}}  \put(54,6){\line(0,-1){6}}
\put(54,12.5){\line(1,0){7}} \put(54,4.5){\line(1,0){7}} 
\put(61,12.5){\line(0,-1){4}} \put(61,4.5){\line(0,1){3}}
\put(59,8.5){\line(1,0){4}} \put(59,7.5){\line(1,0){4}}
\put(64,8.5){\footnotesize Cremat}
\put(64,5.5){\footnotesize CR-112}
\put(57,15.5){\footnotesize veto}
\put(119,15.5){\footnotesize veto}
\put(112,18){\framebox(20,2)}  
\put(83,0){\framebox(10,8){+HV}}
\put(89,10){\framebox(2,4)} \put(90,10){\line(0,-1){2}} \put(90,14){\line(0,1){4}}
\put(92,11){\footnotesize R$_B$}
\put(85,10){\framebox(2,4)} \put(86,10){\line(0,-1){2}} \put(86,14){\line(0,1){2}} \put(86,16){\line(-1,0){18}} \put(68,16){\line(0,1){2}} 
\put(79,11){\footnotesize R$_B$}
\put(71,18){\framebox(40,2)}
\put(105,18){\line(0,-1){3}} 
\put(103,15){\line(1,0){4}} \put(103,14){\line(1,0){4}}
\put(105,14){\line(0,-1){3}} 
\put(100,11){\line(1,0){10}} \put(100,11){\line(1,-1){5}}  \put(110,11){\line(-1,-1){5}}  \put(105,6){\line(0,-1){6}}
\put(105,12.5){\line(1,0){7}} \put(105,4.5){\line(1,0){7}} 
\put(112,12.5){\line(0,-1){4}} \put(112,4.5){\line(0,1){3}}
\put(110,8.5){\line(1,0){4}} \put(110,7.5){\line(1,0){4}}
\put(115,8.5){\footnotesize Cremat}
\put(115,5.5){\footnotesize CR-113}
\put(92.5,15.5){\footnotesize anode}
\put(105,67){\vector(-3,2){50}} \put(105,67){\line(3,0){40}}
\put(40, 100){neutron}
\put(104,65){\circle{4}} \put(104,65){\vector(-3,-1){20}}
\put(106, 60){$^4$He}
\put(97,61){\circle*{1}}  \put(100,65){\circle*{1}} 
\put(98,62){\circle*{1}}  \put(100,62){\circle*{1}}
\put(98,65){\circle*{1}}  \put(90,61){\circle*{1}}
\put(95,63){\circle*{1}}  \put(93,60){\circle*{1}}
\put(88,59){\circle*{1}}  \put(99,63){\circle*{1}}
\end{picture}
\caption{A schematic of the detector: the drift field is created by a cathode mesh, field-shaping rings attached to a resistor chain, and a ground mesh.  Primary ionization from a recoiling nucleus drifts down to the ground mesh. The high-field amplification region is formed by the ground
mesh and the anode plane.  The ground mesh is read  out with a fast amplifier and the veto and anode are readout with charge-sensitive preamplifiers. 
Scintillation light from the amplification region is recorded with the CCD camera.
\label{Detector_Drawing}}
\end{figure}

\begin{figure}
\center
\begin{tabular}{ll}
\includegraphics[height=7.5cm]{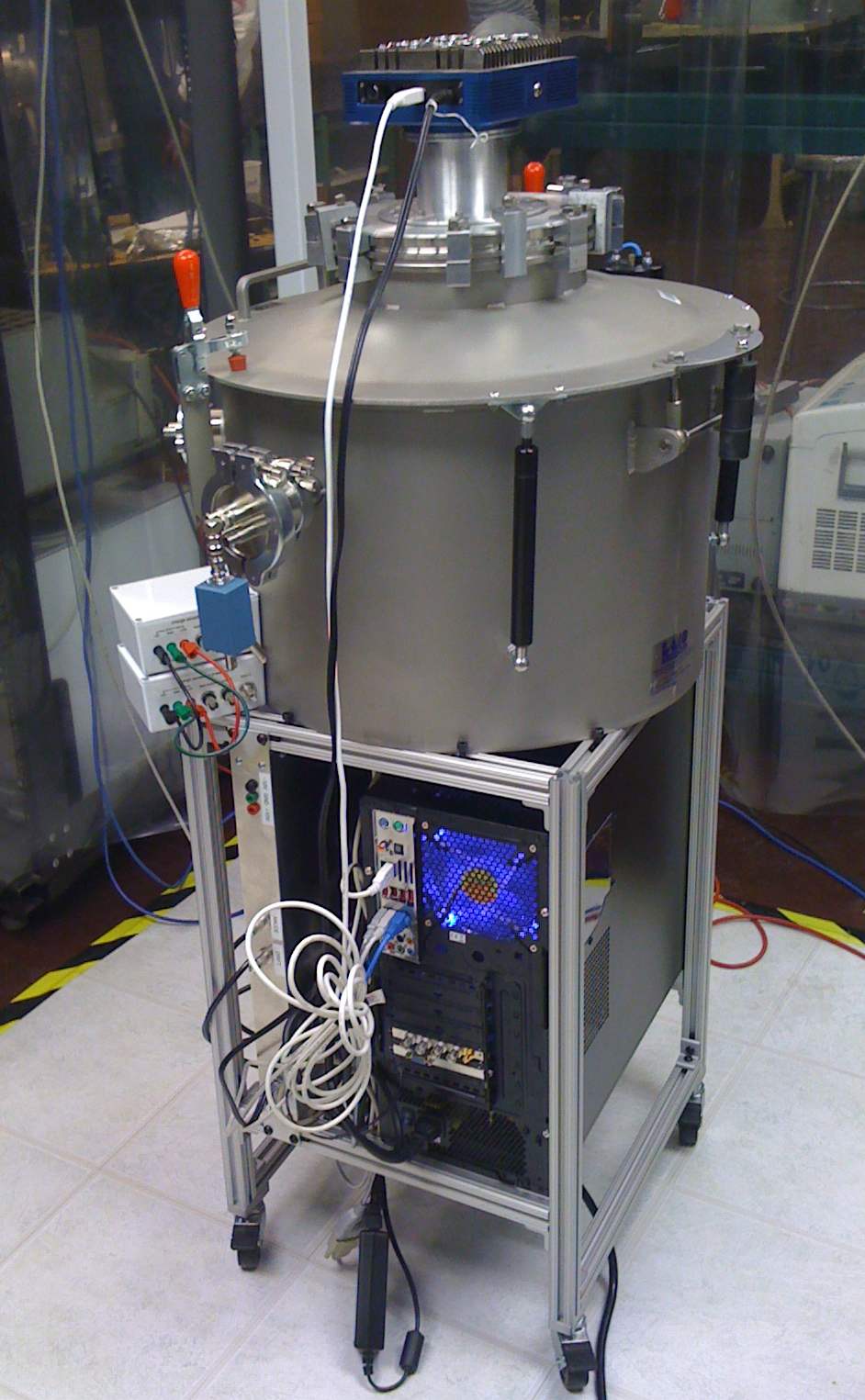}  
& \includegraphics[height=7.5cm]{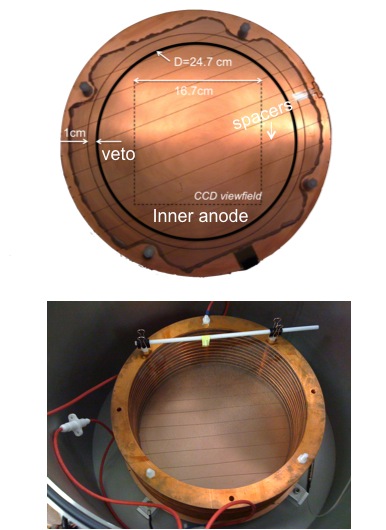} 
\end{tabular}
\caption{ Left: The first generation detector. The detector vessel is sitting on top of the cart which holds the readout electronics. The CCD camera is mounted on the top flange. Right-top: A diagram  
  of the anode plane. Right-bottom: A photograph of the field cage and amplification plane of the TPC. The ground mesh is difficult to see, however the spacers separating it from the anode plane are visible as horizontal lines. }
\label{Detector_Picture}
\end{figure}

We have developed a first generation of DCTPC.
The TPC has a drift volume
above an amplification plane that has two active regions.  The amplification plane is imaged with one CCD, and the charge is read out for each active region. 
The redundancy in the readout of the amplification plane is powerful for event reconstruction and background rejection. 
The detector schematic is shown in the Fig.~\ref{Detector_Drawing} and a photograph of the detector vessel is
shown in Fig.~\ref{Detector_Picture}.

\subsection{The Time Projection Chamber}

The detector follows the classic design of TPCs.  
We have chosen a mixture of CF$_4$ and He as our operating gas.  We study
several gas ratios below, but most data were taken with the gasses
mixed at 75 Torr and 525 Torr respectively.   Helium and fluorine
both have large cross sections for fast neutron scattering
\cite{endf}.  The $^4$He nucleus is closer to
the neutron mass, and therefore the recoils will produce longer tracks
than the fluorine nucleus.

Within the vessel, a uniform drift field is molded by a series of copper rings with an inner diameter of 26.9~cm.  The rings are connected with 
resistors, see Fig.~\ref{Detector_Drawing}. The top ring carries a high transparency stainless-steel mesh with wire pitch of 512~$\mu$m and 
wire diameter of 31~$\mu$m. This top mesh is maintained at a negative
high voltage. 
The total height of the field cage is 10~cm. 
Primary ionization electrons are drifted down to a grounded stainless-steel mesh with wire pitch of 256~$\mu$m and wire diameter of 30~$\mu$m.

The amplification region is created by the ground mesh and the copper anode that are separated 
 with  insulating tubes, called spacers,  440~$\mu$m in diameter, placed 2.56~cm apart~\cite{Denis}.
This high-field amplification gap multiplies the primary  ionization
charge by a factor of approximately $10^5$, as we discuss below. 

The anode bias
scheme is shown in Fig.~\ref{Detector_Drawing}, where resistor $R_{B}$ is 210 M$\Omega$.
The anode of the amplification plane is divided into two regions.  The inner part of the anode, diameter 24.7~cm, is used for the detection of tracks. 
A veto ring, with inner diameter 24.8~cm and outer diameter 26.8~cm, surrounds the central plane and  is used to veto charged particles originating from decays 
from within the detector components. 
A central square region, 16.7~cm~$\times$~16.7~cm, of the amplification plane is read out optically using the CCD camera. 
Using this geometry, we calculate that the active volume of the detector that has a charge readout is 5.6~L, while the volume with both CCD and charge readout is 2.8~L.

\subsection{The Gas Mixture and Associated Operating Voltages}

\begin{table}[tb]
\begin{tabular}{lcccccccc}
\hline 
CF$_4$ & P   & V$_{\rm{anode}}$ & V$_{\rm{drift}}$ & G$_{\rm{CCD}}$    & G$_{\rm{anode}}$ & Gas & Threshold \\
{\small \%}      & {\small [Torr]}     &  \multicolumn{2}{c}{\small [V]}      &  {\small [ADU/keV$_\alpha$]}  &  \multicolumn{1}{c}{\small [mV/keV$_\alpha$]}  &  gain & {\small [mV]}\\
\hline
100 & 75 & 680 & -1200 & 16.3 &0.346  &$1.1\times10^5$  & 20.3\\
12.5  & 600 & 700 & -1600 & 9.18 &0.187  &$>6.0\times10^4$ & 18.8\\
6.25 & 600 & 600 & -1600 & 15.3 &0.518  & $>1.6\times10^5$ &20.3\\
\hline 
\end{tabular}
\caption{Summary of detector calibrations for running in the various
  gas mixtures.    We report the run conditions for the 6.25\% mixture
  here, however due to instability,  we do not report on this mixture further (see text).  The trigger threshold is that used for neutron and background running.  Higher thresholds are needed for the energy calibration runs.
}
\label{table:calib}
\end{table}

In order to select the gas mixture,
we studied three gases: 100\%CF$_4$, 12.5\%
CF$_4$ -- 87.5\% He, and  6.25\%
CF$_4$ -- 93.75\% He.   
Operating voltages, listed in Table~\ref{table:calib}, were optimized
for each gas.   However, we had difficulty finding suitable operating
conditions for the 6.25\% mixture.   Small variations in voltage caused
very large gain differences.  We also found that the absolute and relative 
energy calibrations were inconsistent over several gas fills with this
mixture, likely
due to the much steeper 
increase of gain with respect 
to anode voltage.  
Therefore, we rejected this mixture and do not report the results of the 6.25\% CF$_4$ below.

\subsection{CCD Readout \label{ccd}}

The CCD camera records scintillation light created in the
amplification region, see Fig.~\ref{Detector_Drawing}.  
The camera is the U6 model made by Apogee Instruments Inc., with the Kodak KAF-1001E CCD chip. The chip is kept at $-$20~$^{\circ}$C to  minimize the dark current.  
An image from the amplification plane is focused to the CCD chip using
a Nikon lens, of focal length 55 mm and set to f/1.2.  The CCD chip consists of 1024 by 1024 square
pixels where the side length of a pixel is 24~$\mu$m. The active area
on the anode plate is a square of 16.7~cm $\times$~16.7 cm, see Fig.~\ref{Detector_Picture}.
In order to maximize the signal-to-noise ratio, we merge the CCD
pixels into 4x4 bins 
during readout, so that
the effective spatial resolution is 652~$\mu$m.  The unit for one pixel count returned by
the CCD readout is an ADU (Analog-to-Digital Unit).

The camera does not use a shutter and the CCD sensor is always live,
even during the CCD readout. 
This leads to two types of exposures: a nominal exposure lasting  1
second, and a 
shorter (typically 250 ms) `parasitic'
exposure taken during the CCD readout and event processing. 
Track images recorded during the short parasitic exposure can be shifted in the direction of
readout of CCD rows, and do not have corresponding signal in the charge readout channels. Therefore, they are easy to remove during analysis.

\subsection{Charge Readout}

The charge readout is very powerful in suppressing CCD-specific
backgrounds ({\it e.g.,} direct hits in CCD chip and residual bulk images).
This is the major improvement over the CCD-only approach described in Ref.~\cite{Roccaro:2009tg}. The anode plate contains three regions separated by electrically
insulated gaps, as shown in Fig~\ref{Detector_Picture}. The outermost
circular region is connected to a common ground, onto which the ground
mesh is attached. The inner disk region is the actual anode,
maintained at 680~V.  The circular-ring region between is used as a
veto and is kept at the same potential as the anode. 

When a particle enters from outside, its path must cross the veto
region.  Therefore, we can detect this type of event by inspecting the charge
signal from the veto region.  The veto is not applied at
the trigger level. The veto signal is stored for use in the analysis.

The anode signal is collected using a charge-sensitive preamplifier (CSP), the Cremat CR-113 with a gain of 1.5~mV/pC. The pulse height of the output of the CSP corresponds to the total charge collected on the anode. Therefore, the pulse height is used to reconstruct the total energy of the event. The veto electrode is read out through the more sensitive CSP, the Cremat CR-112 with a gain of 15~mV/pC. The ground mesh is read out through a fast current amplifier, the Route2Electronics HS-AMP-CF-2nF with 1~ns rise time and gain of 80. The fast amplifier has a built-in protection that guards against discharges from this high-capacitance detector. The CSPs are protected with 300~$\Omega$ resistors.  All three signals are digitized using the ATS860, a 250~MS/s 8-bit flash ADC made by AlazarTech. 
The raw unit for charge readout in the the following discussion is mV.  The digitization of all three channels is triggered by a signal on the anode and stored with the current CCD image. The trigger threshold used is gas dependent.

\subsection{Event Readout, Reconstruction and Correlation \label{correlate}}

The combination of the four measurements, CCD image, charge deposited on the inner anode and veto anode and the current recorded at the ground mesh, is used for reconstruction. The CCD track reconstruction starts with subtracting the bias image
that is obtained at the beginning of every run. At this time, any
unusually high-valued pixels are removed.  The image is then
smoothed. With the smoothed image,  a search is conducted for clusters
of adjacent pixels with values 3.7$\sigma$ larger than the mean value
of the image \cite{DMTPC}.  If enough such pixels are found, the cluster is determined to be a possible track.  An additional ring of pixels around the main cluster is added to account for diffusion, and the track is the corresponding pixels from the original un-smoothed image.  For each of these tracks, the energy, range, angle, maximum value, and several moments are calculated.

We require that the track be contained within the TPC.  The veto
removes signals from radiation emitted from 
the drift rings or the vessel wall.  However, this also vetoes long tracks produced within
the active volume that enter the veto. Additional inefficiency comes
from the requirement 
that the track be fully contained in the CCD view-field.   
We show below that these are not a significant constraints, even in
the first generation detector.

Tracks imaged by the CCD camera are matched to charge waveforms using
the reconstructed energy.   The rate of events in the detector, even
with the neutron and $\alpha$ sources, is relatively low, and most
exposures contain no tracks. 

\section{Calibration}

\begin{figure}[h]
\begin{center}
\includegraphics[scale=0.20]{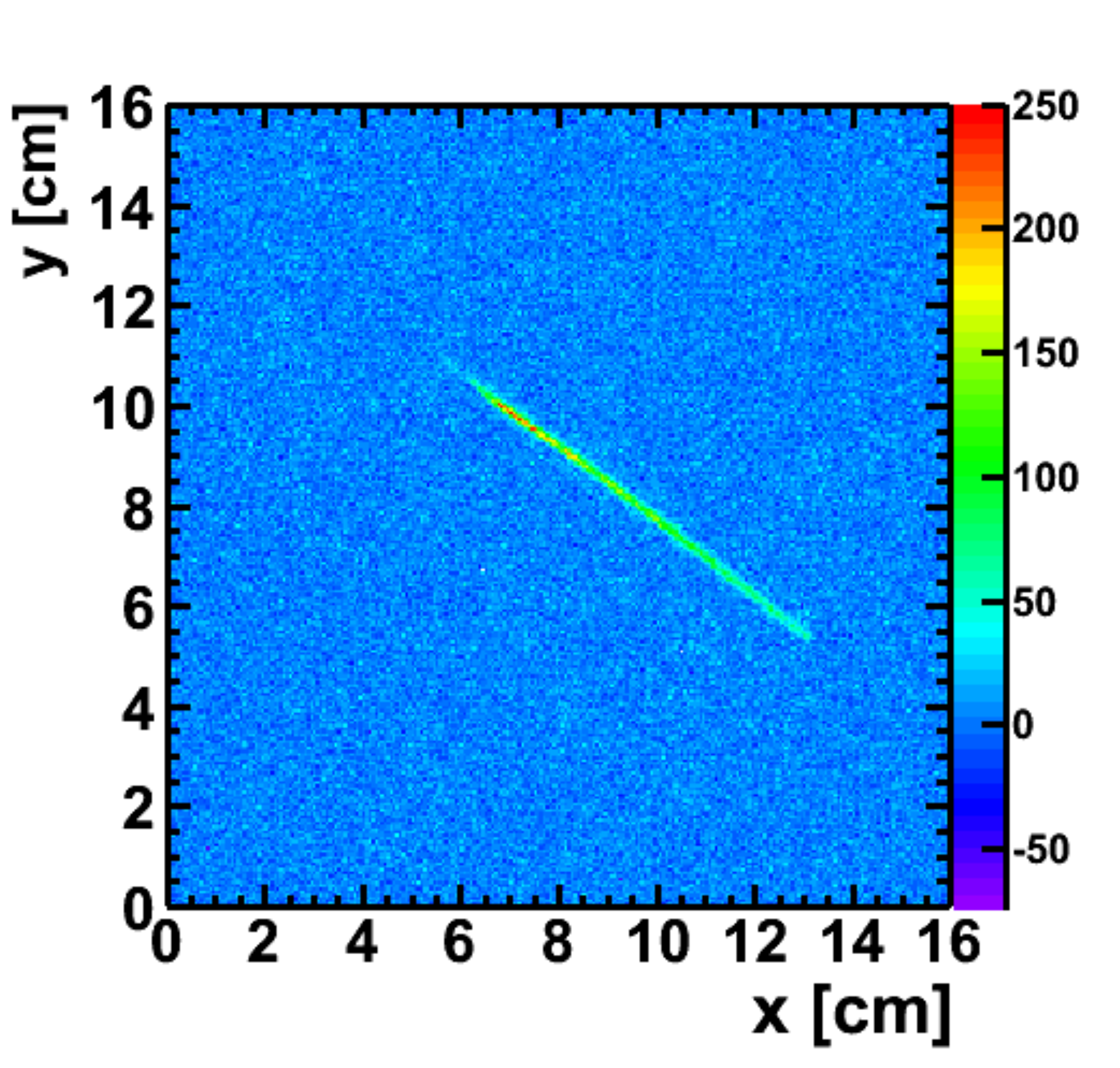}
\includegraphics[scale=0.20]{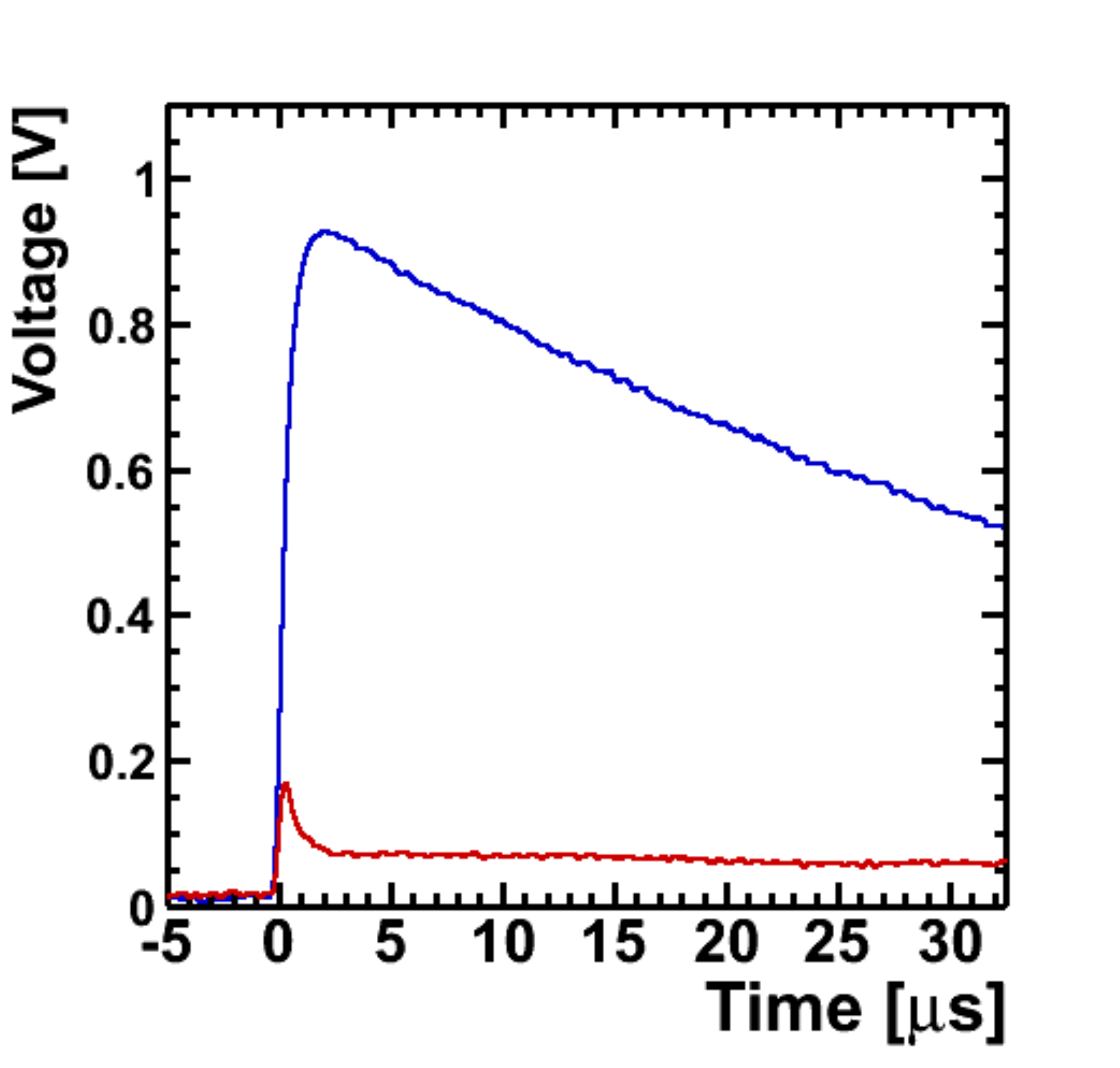}ø
\includegraphics[scale=0.20]{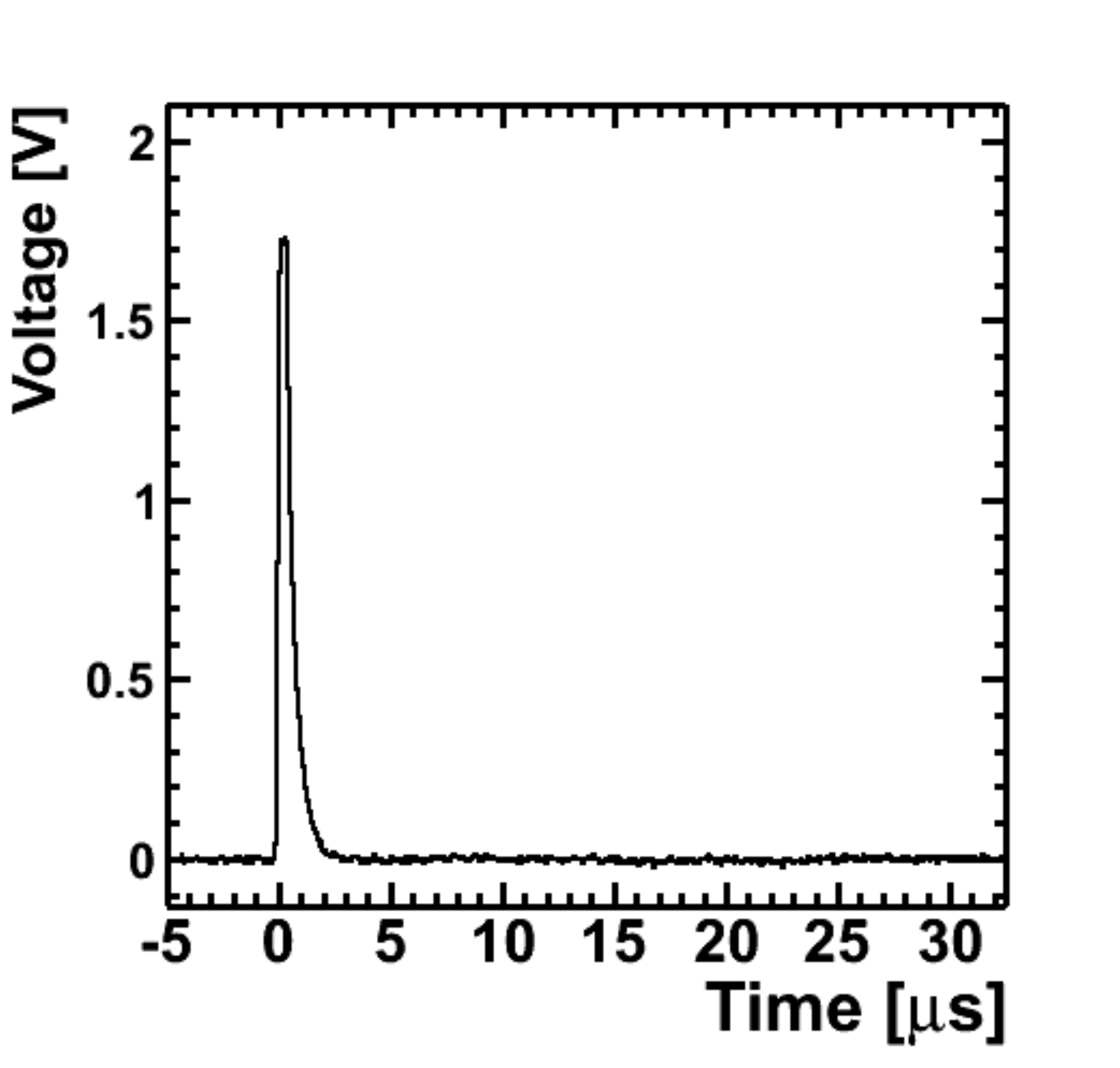}
\\
\includegraphics[scale=0.20]{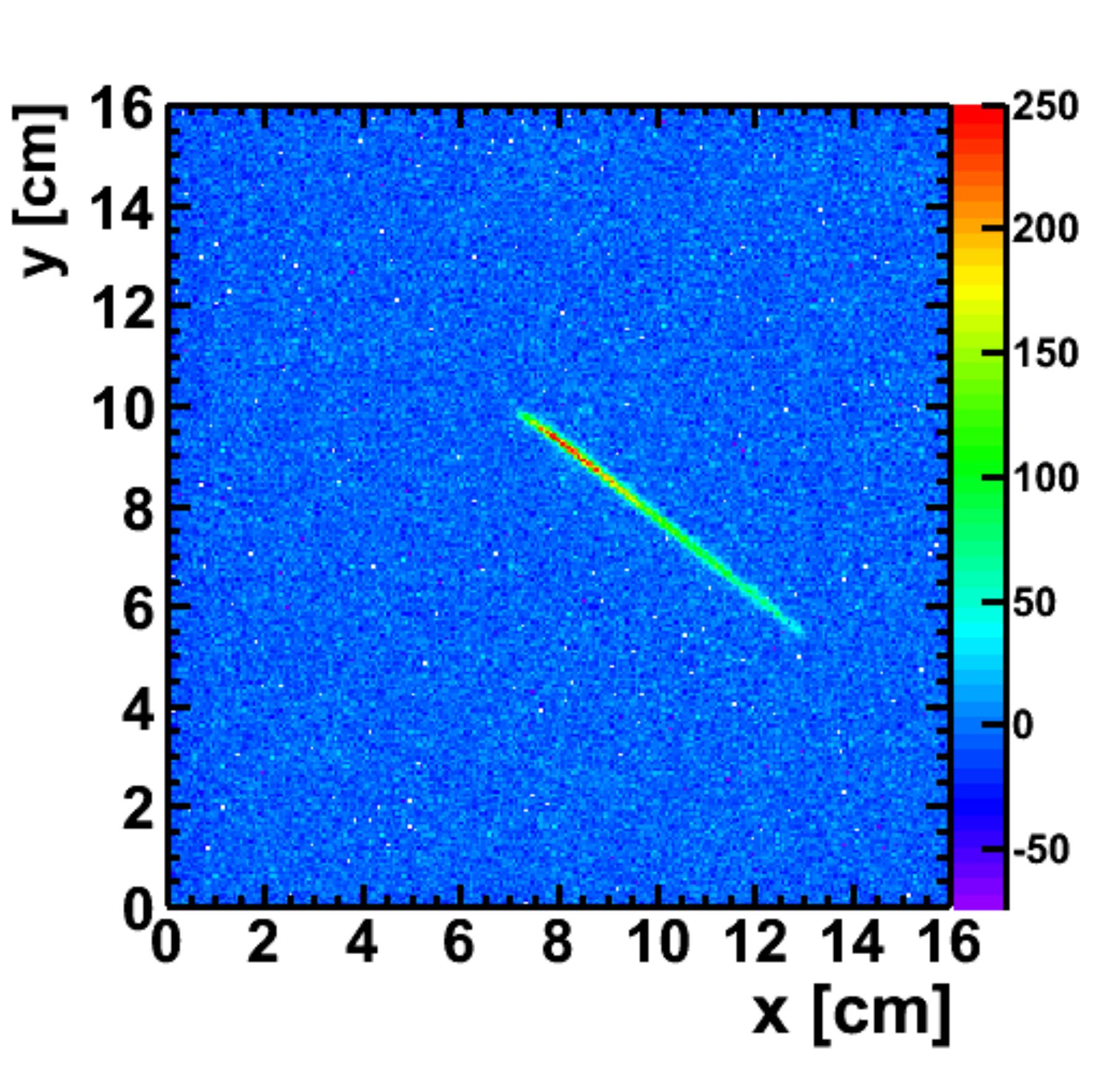}
\includegraphics[scale=0.20]{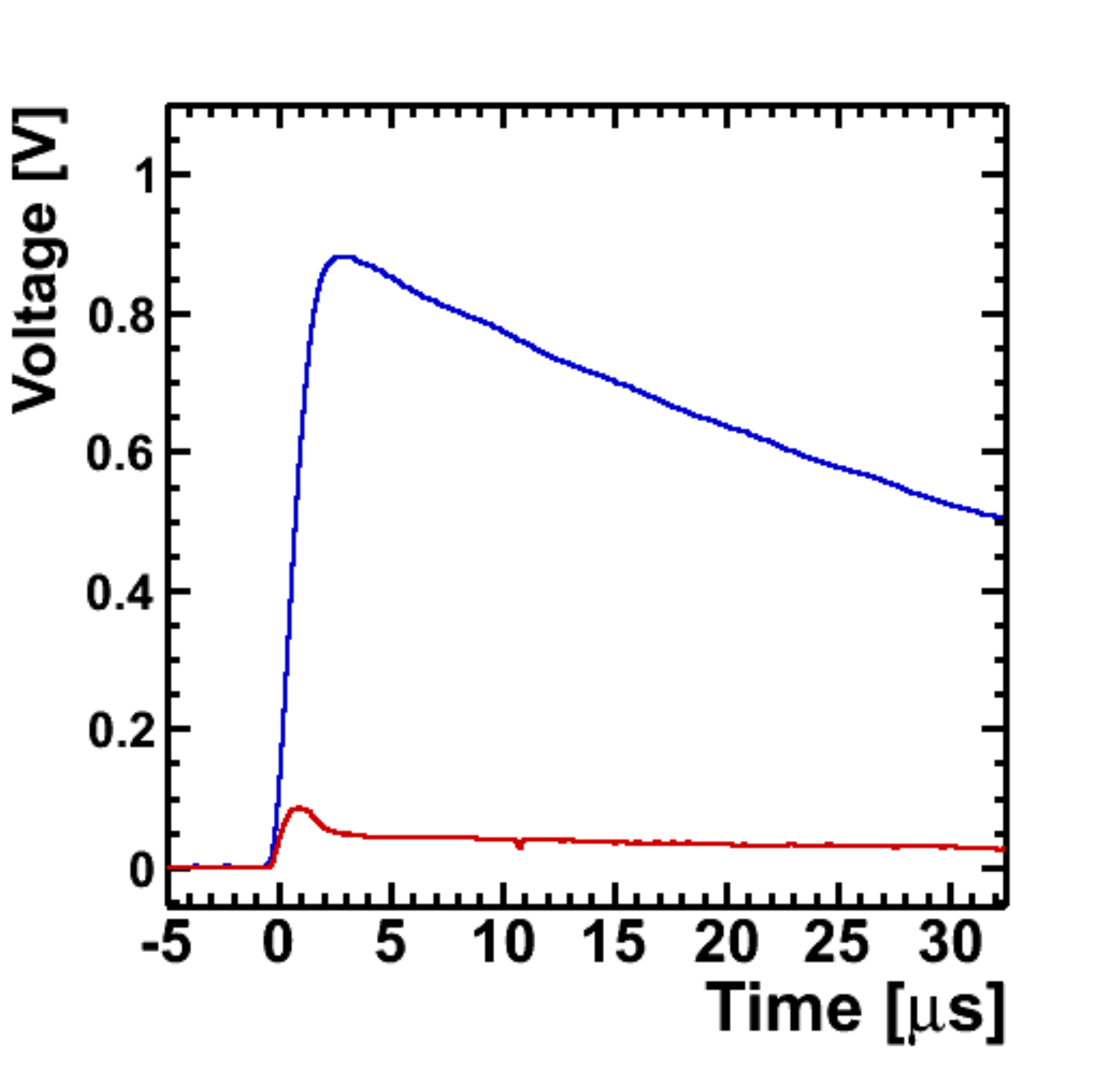}
\includegraphics[scale=0.20]{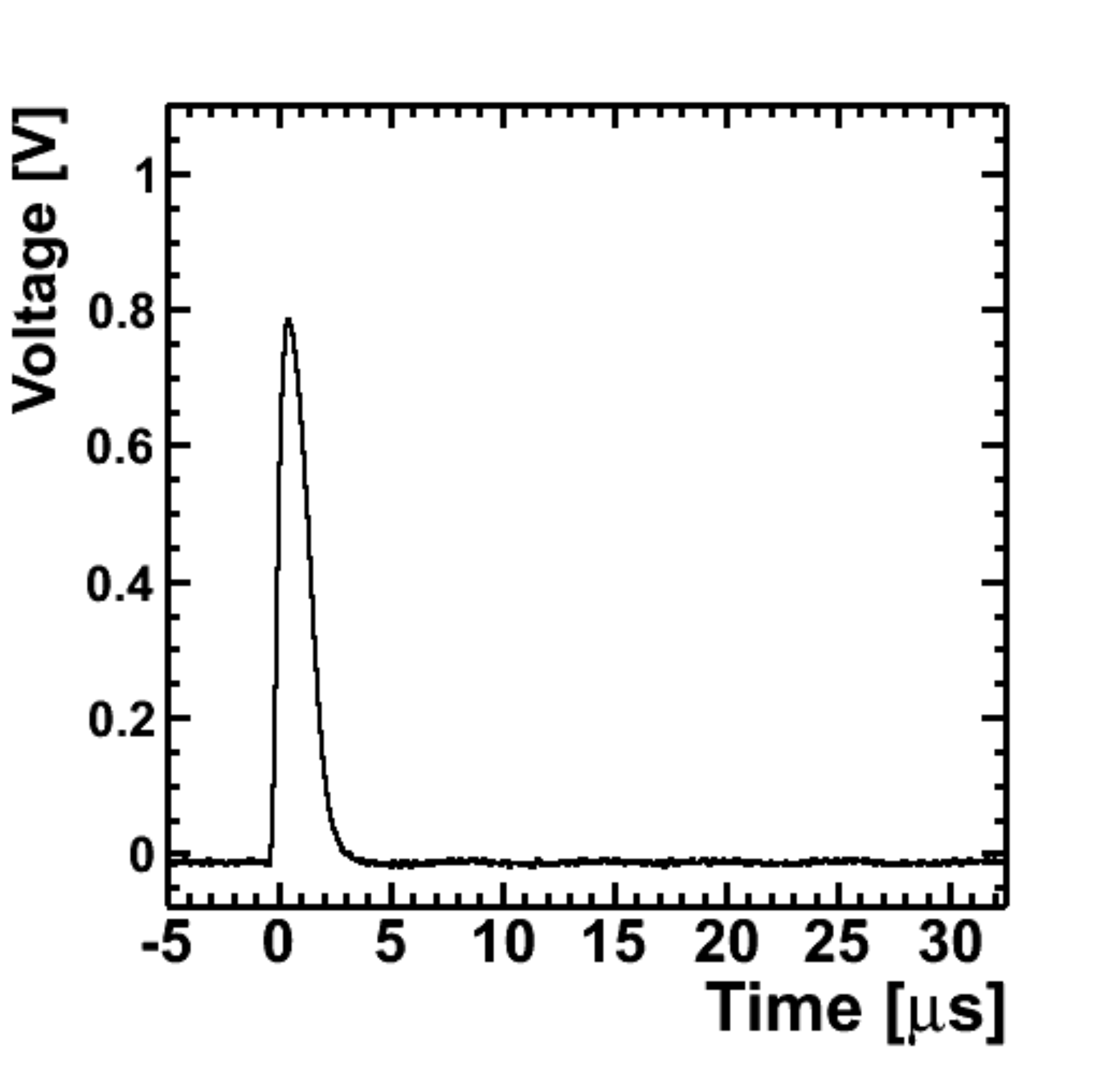}
\caption{
	For $\alpha$ particle tracks from the $^{241}$Am source,
	images of the CCD and charge read-out system are shown.
	Left: CCD image of $\alpha$ particle tracks. 
	Middle: Waveforms from the Veto and Anode charge read-out. 
	Right: Waveforms from the Ground-Mesh charge read-out.
	Top: 100\% CF$_{4}$;
	Bottom: 12.5\% CF$_{4}$ + Helium.
	\label{exampleAlpha}
}	
\end{center}
\end{figure}

Calibration of the detector was performed using a $^{241}$Am $\alpha$ source. A window on the source degrades the energy of the $\alpha$ particles leading to a mean energy of $4.40\pm0.04$~MeV \cite{HaykThesis}. Example $\alpha$ events in two different gas mixtures are shown in Fig.~\ref{exampleAlpha}.

\subsection{Spatial Calibration}

The spatial calibration is obtained by measuring the positions of the wire spacers separating the anode 
plane from the ground mesh, shown in Fig.~\ref{Detector_Picture}.  The spacers create a region of 
low gain. A $^{137}$Cs source is used to generate electronic recoils so that a small but diffuse amount of light is seen in each image.
By integrating many images, the spacers become visible as regions with low light levels. The spacers are placed 2.5 cm apart, and the fit gives a calibration of 163~$\mu$m per pixel.  The total area of the 1024 by 1024 pixel region imaged by the CCD is then
 16.7~cm $\times$ 16.7~cm, as discussed in Sec.~\ref{ccd}.  The uncertainty on the spatial resolution
 is believed to be dominated by systematics from misplacement of spacers, optical distortions, and deviations
 in the spacer shape from a line.

\subsection{Energy Calibration}

%
%

The total ionization left by the track is measured in three ways:
(1) integrating the light in the CCD track,
(2) measuring the pulse peak height of the central anode signal and
(3) integrating the current signal from the mesh channel.
At the drift voltage used in these data, the $\alpha$ tracks saturate
the digitizer of the ground mesh and possibly the anode preamp.   Using an 8-bit ADC, it is not possible to have both a fine
energy  granularity and a large range.
The gain is optimized to have good energy resolution for low-momentum
recoils created in the neutron scattering rather than the $\alpha$
studies. For this reason only methods (1) and (2) are used for
determining track energy.   Electron attachment was measured to be negligible at our drift
fields  \cite{Caldwell:2009si}.

\begin{figure}[t]
\begin{center}
\includegraphics[scale=0.2]{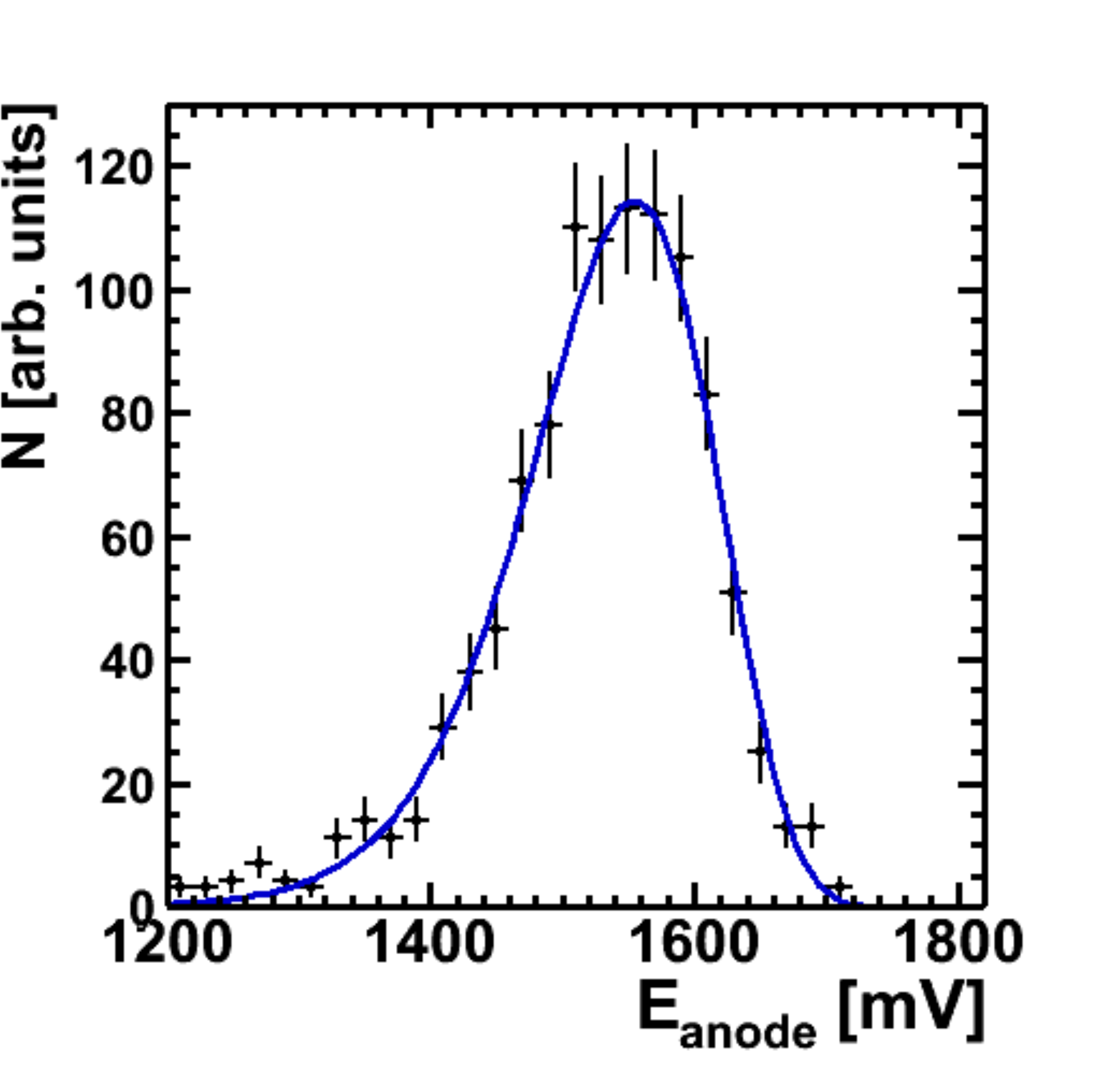}
\includegraphics[scale=0.2]{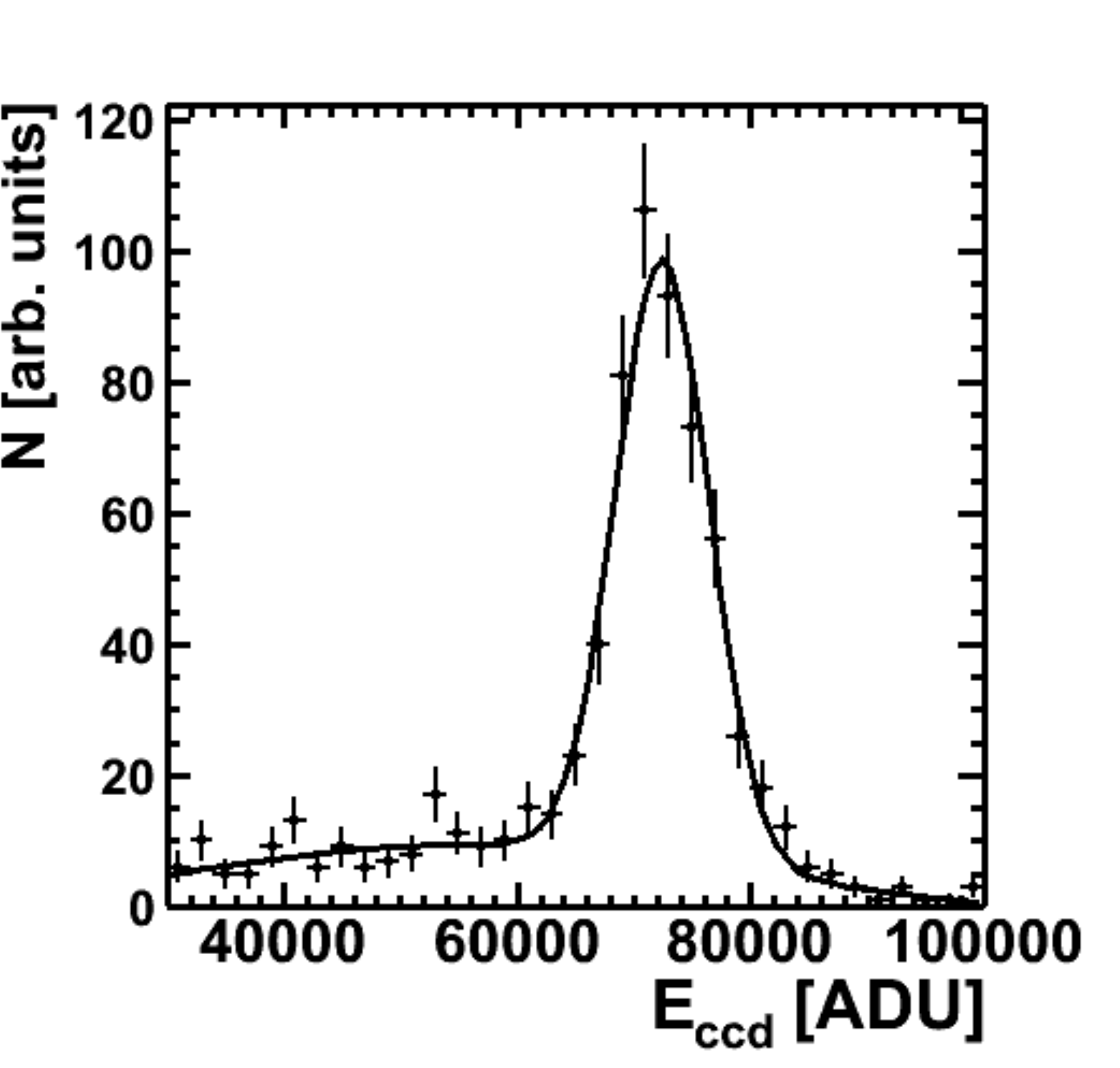}\\
\includegraphics[scale=0.2]{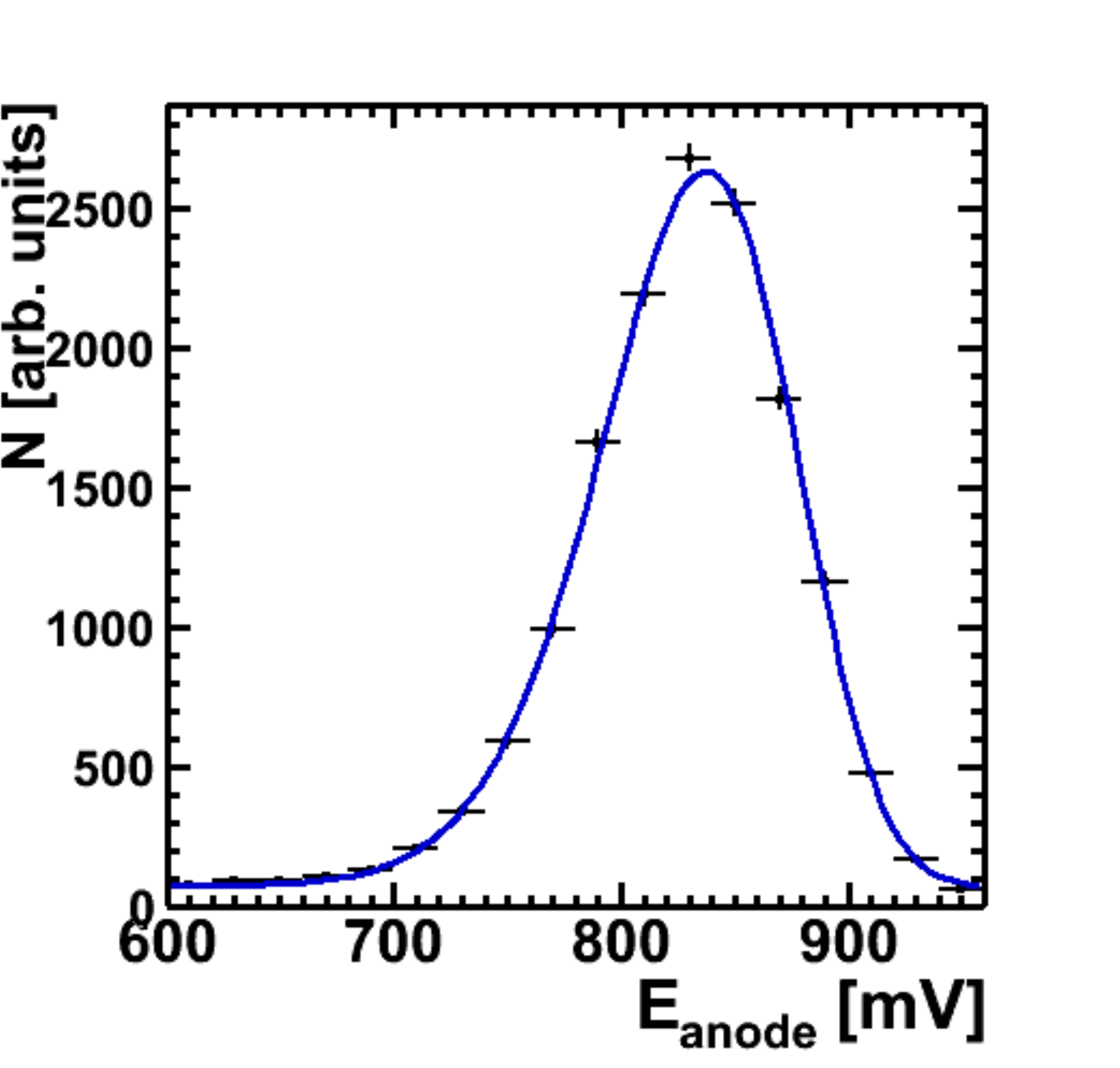}
\includegraphics[scale=0.2]{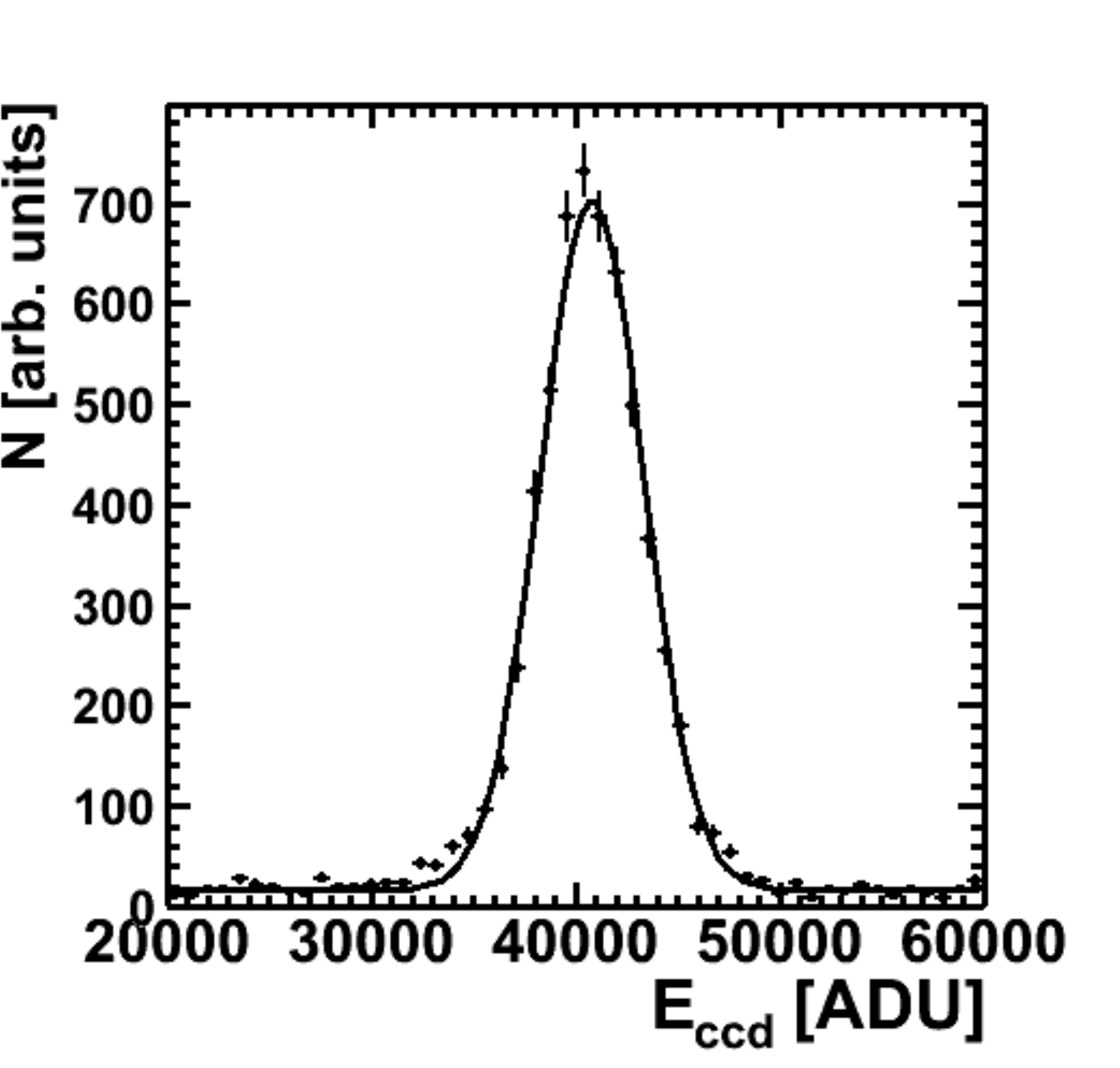}
\caption{Energy gain calibration using the charge read-out (left) and CCD (right).
	for 100\% (top) and 12.5\%, (bottom) CF$_{4}$ mixtures. 
}
\label{Energy_Calib}
\end{center}
\end{figure}

\begin{figure}
\begin{center}
\includegraphics[scale=0.30]{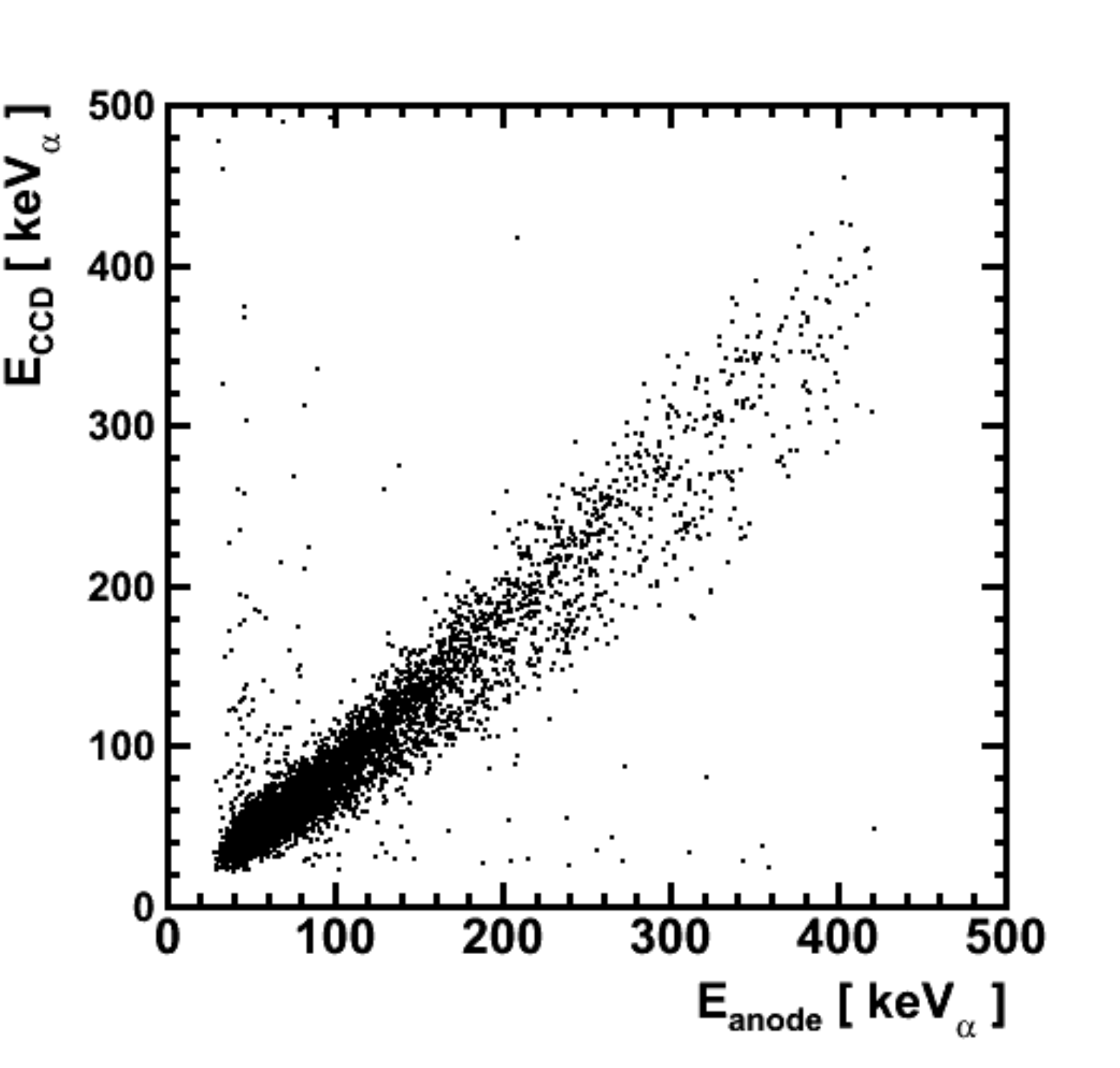}
\includegraphics[scale=0.30]{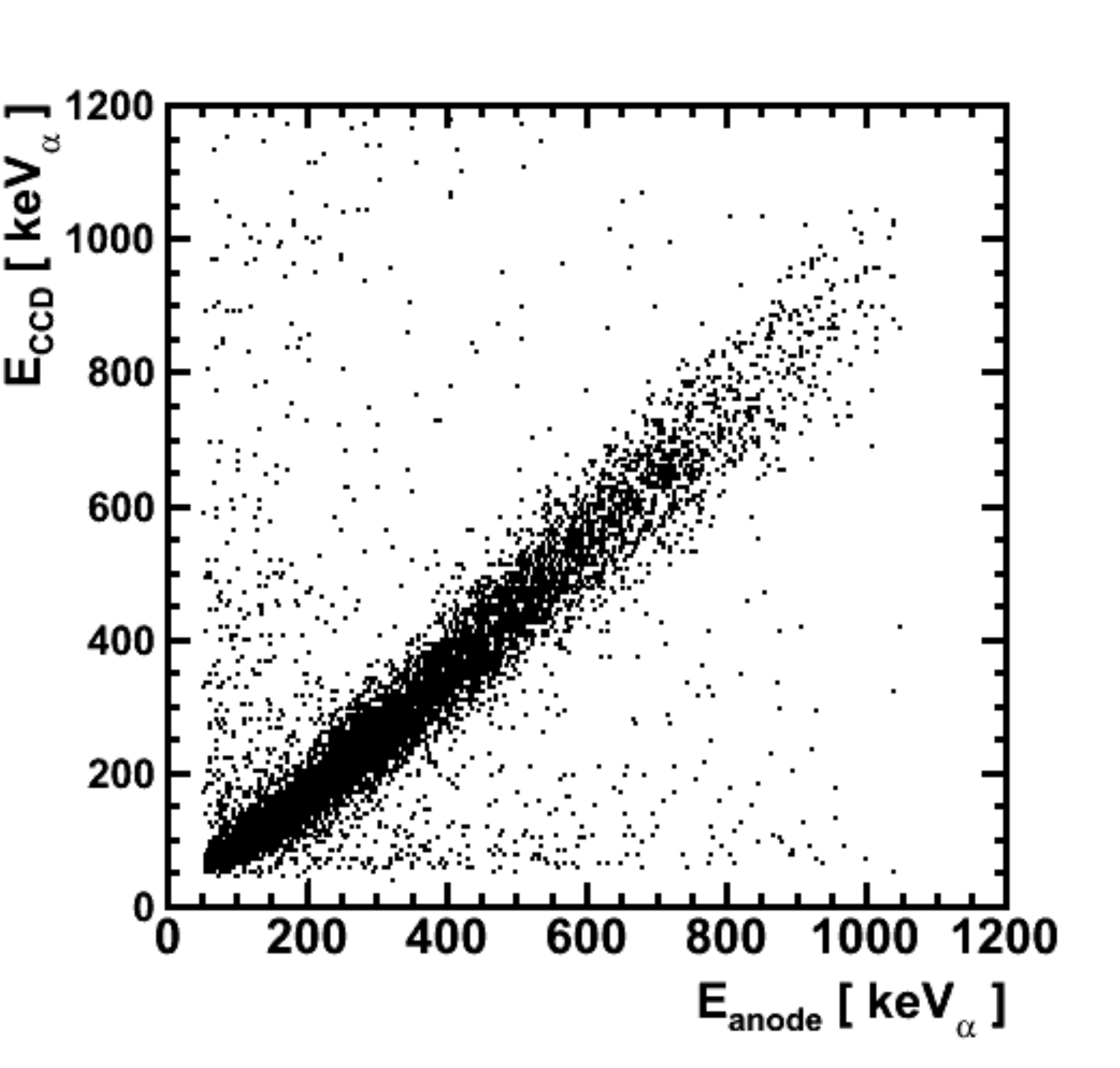}
\caption{
	To validate the $\alpha$ source calibration, we use CCD light versus charge read-out signal for neutron 
	scatters from a $^{252}$Cf source.
	Left: 100\% CF$_{4}$. Right: 12.5\% CF$_{4}$ + He. 
	\label{fg::AnodeCountsCorr}
}
\end{center}
\end{figure}

The energy distributions from the anode and CCD are shown for two gas mixtures in
Fig.~\ref{Energy_Calib}.  We have fit the peak position for the energy calibration.
The results of the calibration are given in Table~\ref{table:calib}. 
The energy is quoted in units of alpha equivalent energy to denote that the detector response is calibrated with an alpha source.
We do not apply corrections due to ionization quenching at lower recoil momenta due to lack of experimental data.

The gas gain shown in Table~\ref{table:calib} is calculated from the charge energy calibration, the preamplifier gain of 
1.5 mV / pC, and the gas work function.  The work function of CF$_4$ gas was 
measured to be 33.8~eV \cite{IanThesis}.  This is used for calculating
the gain of all mixtures.

\section{The Neutron Source Run}
\label{sc::recon}

The response of the detector to neutrons was characterized using a
1.24~mCi $^{252}$Cf source.  $^{252}$Cf decays via spontaneous fission
3\% of the time emitting multiple neutrons with a mean energy of
2.35\,MeV \cite{pdg}.    Given the activity of this source, we expect $5.7\times10^6$ neutrons per second in all directions.
The source is placed in a collimator made of borated plastic,  2.1~m from the detector.  
The high flux of both neutrons and x-rays causes an increase in the
rate of sparks in the TPC from $\sim$0.01~Hz to 0.1Hz.  A spark is
identified as an image with a mean number of ADU 1\% greater than the previous image.  Events within 6 exposures of a spark are excluded from analysis to allow sufficient time for the high voltage to recover.

\subsection{CCD, Charge and Matching Cuts }

\begin{table}[tb]
\begin{tabular}{l|c}
\hline 
\hline 
\% of CF$_4$ &  Requirement \\ \hline 
All cases &  Two adjacent pixels with counts $>$3.7$\sigma$ above mean
\cite{DMTPC};\\
& Tracks are contained within $24 \le x,y \le 1000$ pixels; \\
& Maximum pixel $<$ 250 ADU and $<$ 25\% of total signal; \\
& Only one track in the image; \\
& No two tracks within 12 pixels in 1 run \\  \hline
100\% & Tracks shorter than 80 pixels (1.3 cm) \\ \hline
12.5\% & Tracks shorter than 160 pixels (2.6 cm) \\ 
\hline
\hline 
\end{tabular}
\caption{Cuts applied to CCD images for the study.
\label{table:CCDcuts}}
\end{table}

The cuts to isolate good events are applied to the CCD image, the
charge-readout signal and the charge-light matching.
For each run condition described in Table~\ref{table:calib}, we identified a set of optimized cuts. Table~\ref{table:CCDcuts} describes the cuts placed on CCD images.
Only cuts on track-length are gas-specific.  The other cuts are
designed to restrict tracks to the fiducial volume and remove
backgrounds from ionization within the CCD chip, noise artifacts, and
hot pixels.

The charge-readout cuts are more gas dependent than the CCD cuts.  They can be divided into rise time cuts and pulse height cuts, summarized  in Tables~\ref{table:risecuts} and \ref{table:shapecuts} respectively.
For the veto we define two variables: the time is takes the
pulse to rise from 10\% to 90\% of the peak pulse height, $t_{veto}$,
and the ratio of the mesh peak pulse height to the veto peak pulse
height, $R_{veto}$.  For the mesh and anode, we define pulse rise times,
$t_{mesh}$ and $t_{anode}$.  The pulse height range over which $t_{mesh}$ and  $t_{anode}$ are
defined is tuned for each gas mixture. For the mesh we also define two ratios
for the pulse height: $R_{elec}$ the ratio of the electron peak mesh
pulse height to the height of the anode pulse and $R_{ion}$ the ratio
of the ion peak mesh pulse height to the anode pulse height.  $R_{mesh}$
is a simple ratio of the anode pulse height to the mesh pulse height. 

\begin{table}[tb]
\begin{tabular}{l|c|c|c}
\hline 
\hline 
\% of CF$_4$ & $t_{veto}$ & $t_{anode}$  and  $t_{mesh}$   & Pulseheight Range\\ \hline
100\% & $<$800~ns & $t_{mesh}<30$~ns for pulseheight $<100$~mV &  25\% to 75\%  \\
           &                  & $t_{mesh}<25$~ns for pulseheight $>100$~mV & 25\% to 75\%  \\ \hline
12.5\% & $<$800~ns & $t_{anode}<1250$~ns & 10\% to 90\% \\  
\hline
\hline 
\end{tabular}
\caption{Rise time requirements.  Variables are defined in the text. Events are accepted if
  they pass the requirements listed.   The pulseheight
  range over which $t_{veto}$ is measured is always 10\% to 90\%.
  The pulseheight range for the definition of $t_{mesh}$ and $t_{anode}$ vary as listed in column 4.
\label{table:risecuts}}
\end{table}

\begin{table}[tb]
\begin{tabular}{l|c|c|c|c}
\hline  
\hline  
\% of CF$_4$ & $R_{veto}$ & $R_{elec}$ & $R_{ion}$ & $R_{mesh}$\\ \hline
100\% & $>$4 & $1.7<R_{elec}<2.7$ & $1.9<R_{ion}<3.5$  & N/A\\
12.5\% & $>$3 & N/A & N/A &  $1.3<R_{mesh}<3$\\
\hline
\hline 
\end{tabular}
\caption{Pulse height ratio requirements. The ratios are defined in
  the text.    Events are accepted if
  they pass the requirements listed in this table.   
  N/A indicates that this variable is not used. 
\label{table:shapecuts}}
\end{table}

\begin{figure}
\begin{center}
\includegraphics[scale=0.30]{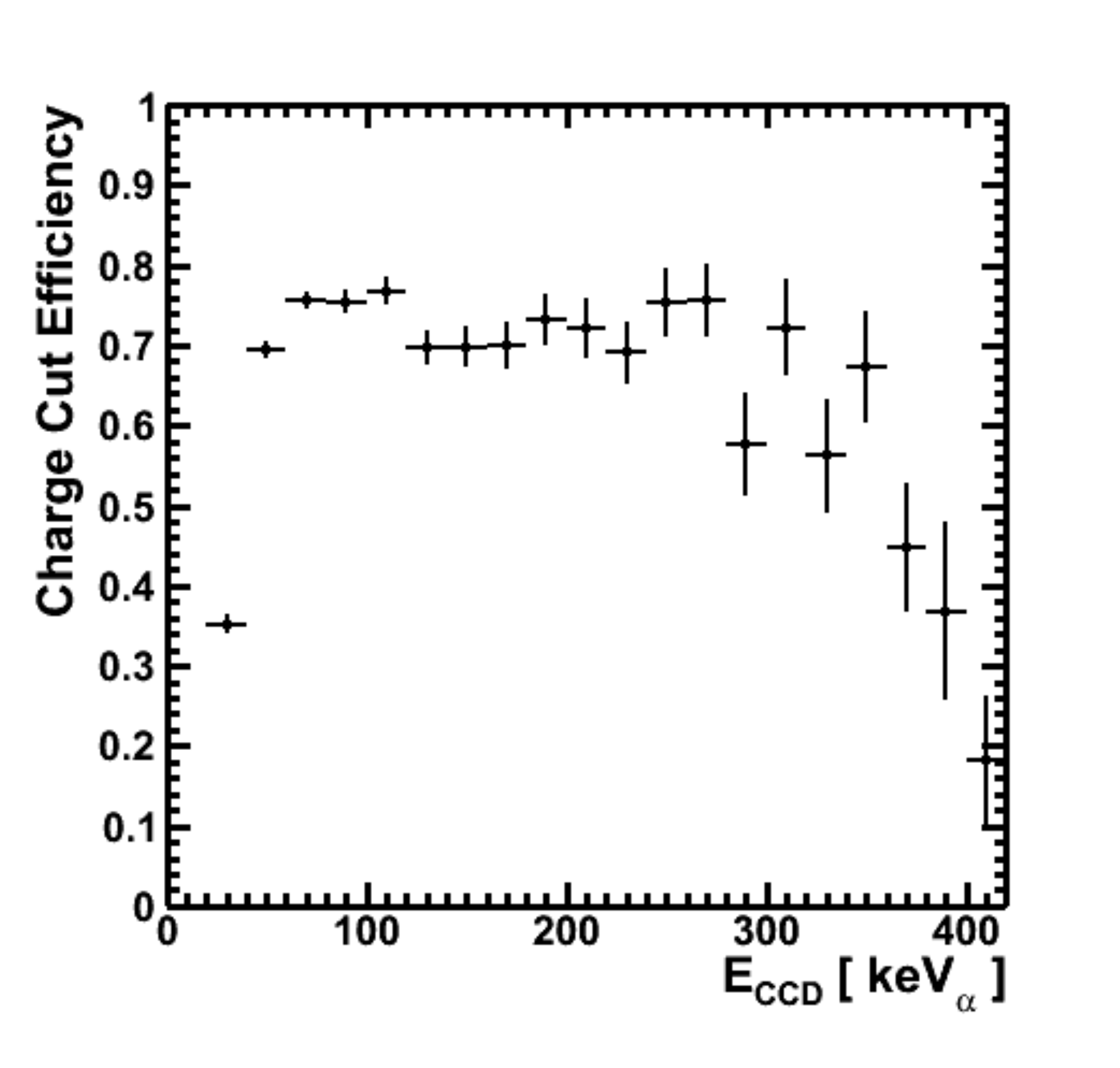}
\includegraphics[scale=0.30]{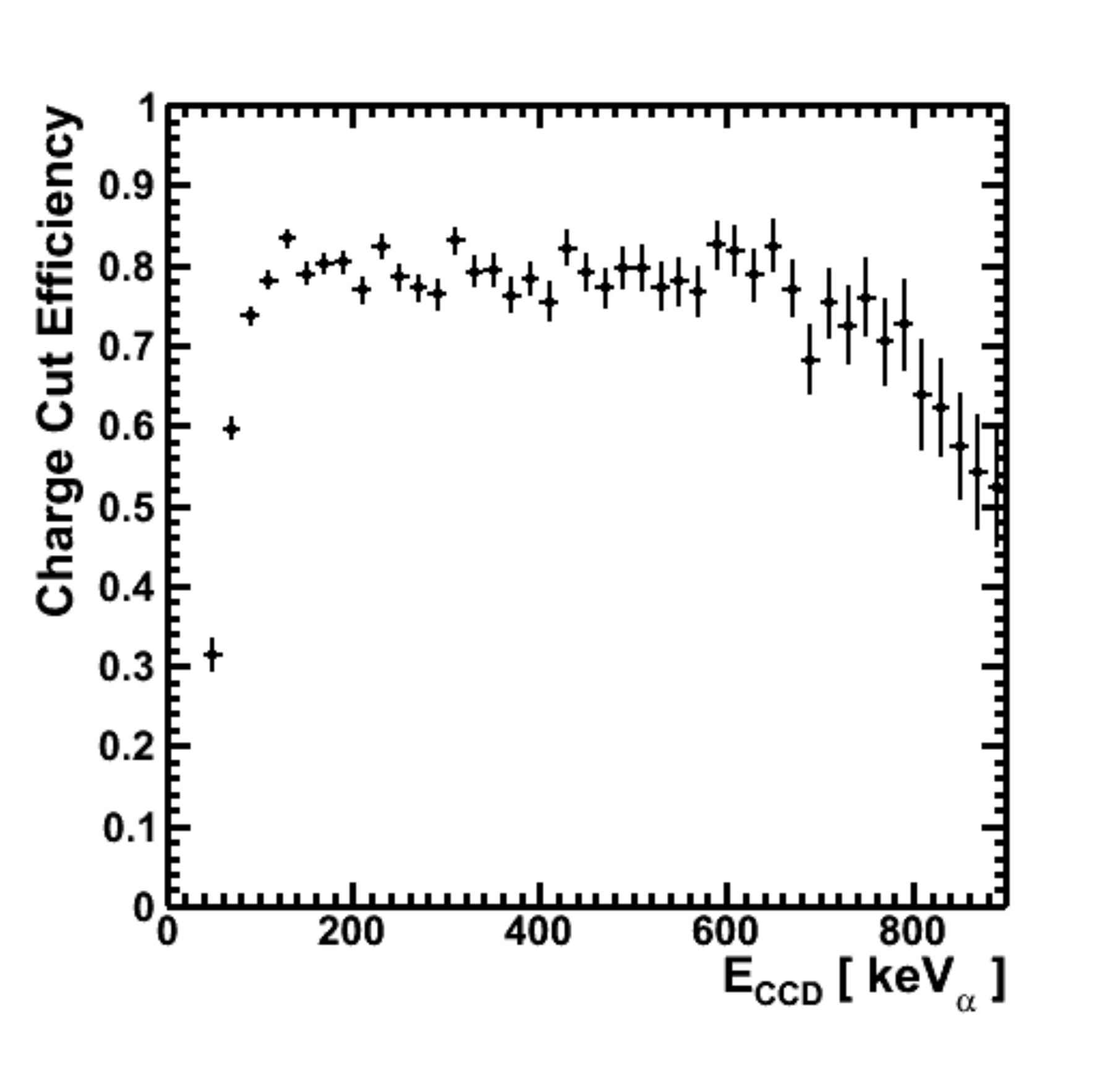}
\caption{
	Efficiency of charge analysis in accepting tracks identified by the CCD analysis.
        The efficiency plateaus at 70-80\% are due to exposure difference between the charge
        and light readout and depend on the event readout time.  The decrease in efficiency at
        higher energies is from the mesh pulse reaching the maximum value read out by the digitizer.
	Left: 100\% CF$_4$ gas. Right: 12.5\% CF$_4$ + He mixture.    The
        errors on the data points reflect the statistics of the calibration
        run.  \label{fig:chargeEff}
}
\end{center}
\end{figure}

The CCD and charge readout run autonomously.  Matching the signals
proceeds by comparing the CCD track energy to the charge energy, where 
the best energy match is identified as the true signal from the track. 
The requirements for matching are shown in Table~\ref{match}.
Optical effects are estimated by measuring the mean ratio of the energies of the light and charge signals in neutron calibration running as a function of the distance from the image center. 
This is used to correct the energy of the light signal to obtain a linear relationship between the two energy measurements. 
The energy from the anode channel is more accurate than the CCD energy, so it is used as the final track energy. Neutron source data shown in  Fig.~\ref{fg::AnodeCountsCorr} confirms a linear relationship between the summed anode energy and the CCD energy, demonstrating a good match between a CCD track and an anode waveform can be found for most of the events.

\begin{table}[tb]
\begin{tabular}{l|c}
\hline 
\hline 
\% of CF$_4$ &  Requirement \\ \hline 
100\% & $|\Delta E| < 40$ keV$_\alpha$  if $E_{anode}<150$
keV$_\alpha$;\\
           & $|\Delta E|< 75$ keV$_\alpha$  if $E_{anode}>150$
keV$_\alpha$;\\ \hline
12.5\% & $|\Delta E| < 75$ keV$_\alpha$ for $E_{anode}<250$ 
keV$_\alpha$; \\
& $|\Delta E|<125$ keV$_\alpha$ for $250<E_{anode}<500 $ keV$_\alpha$;\\
& $|\Delta E|<150$ keV$_\alpha$ for $E_{anode}>500$ keV$_\alpha$ \\ \hline
\hline
\hline 
\end{tabular}
\caption{Requirements for CCD and charge event matching in this study, where
$|\Delta E|=|E_{CCD}-E_{anode}|$.
\label{match}}
\end{table}

The efficiency of the cuts, including charge-light matching for
neutron events is given in Fig.~\ref{fig:chargeEff}. This data is dominated by nuclear recoils rather than CCD artifacts, so is ideal for the efficiency determination.  The efficiency is defined as the ratio of the number of tracks passing all CCD cuts, charge cuts and charge-light matching compared to the number of CCD tracks found without using the charge signals.

\subsection{Discussion of Neutron Source Run}

\begin{figure}
\begin{center}
\includegraphics[scale=0.20]{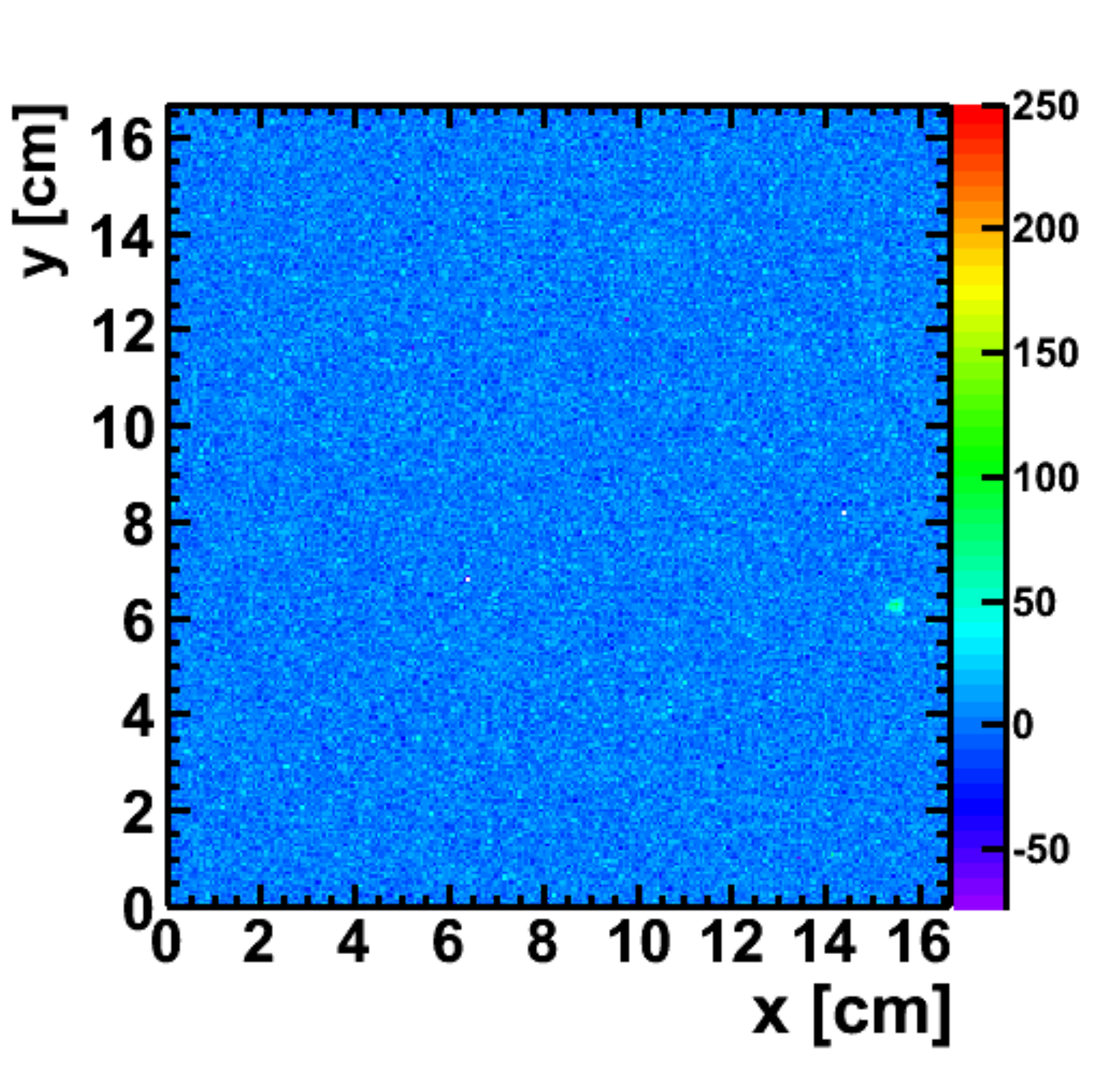}
\includegraphics[scale=0.20]{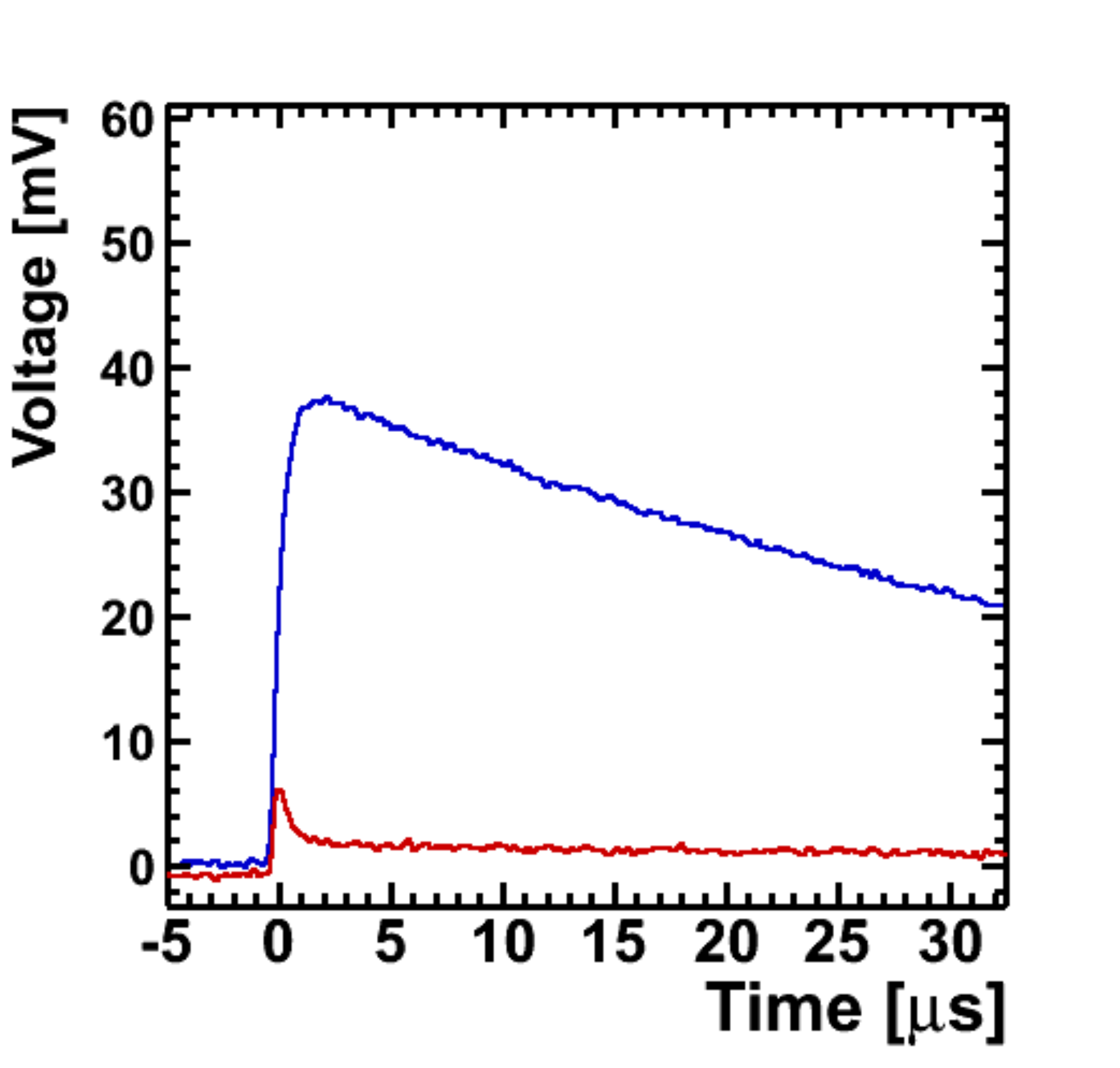}
\includegraphics[scale=0.20]{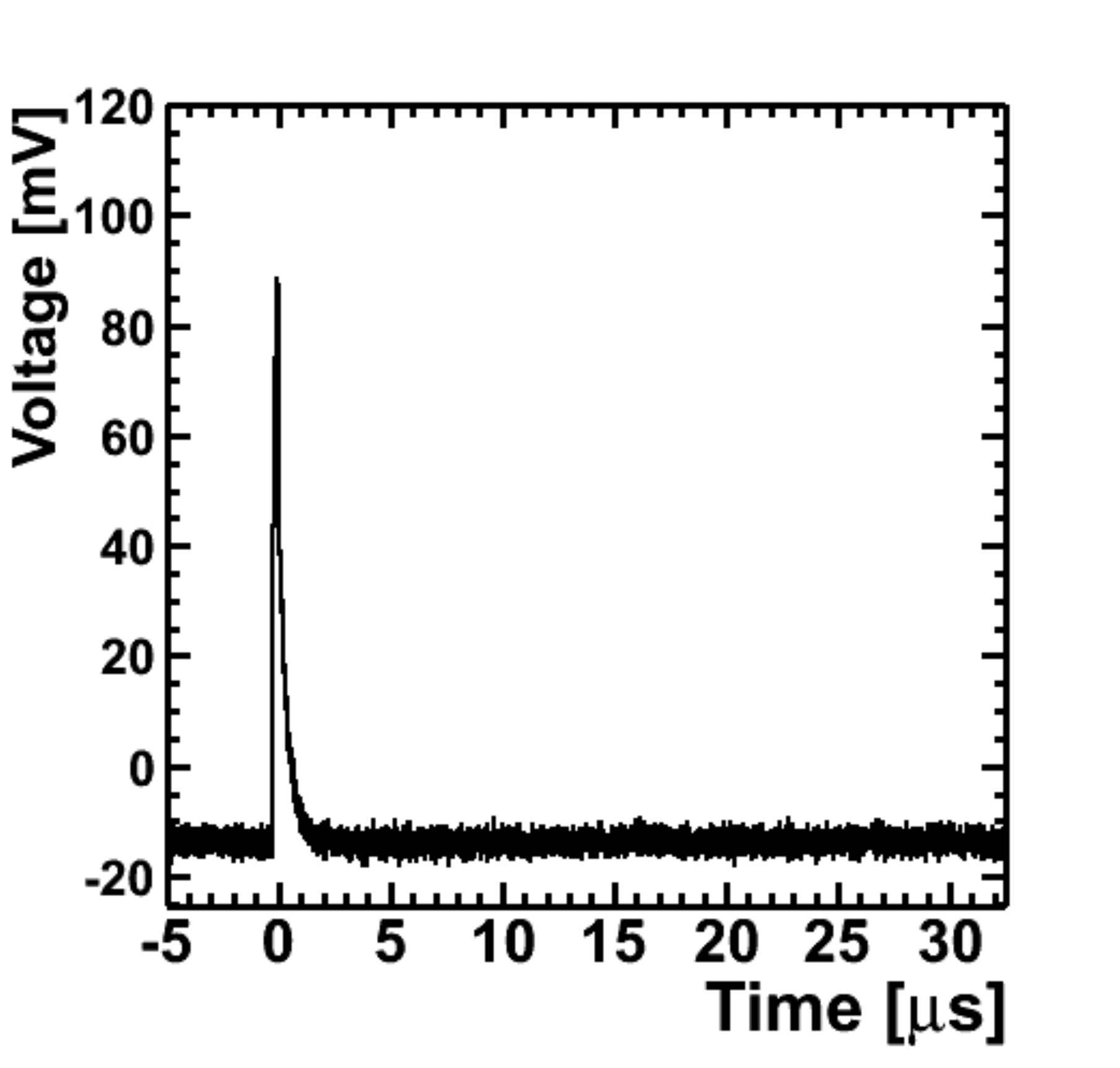}
\\
\includegraphics[scale=0.20]{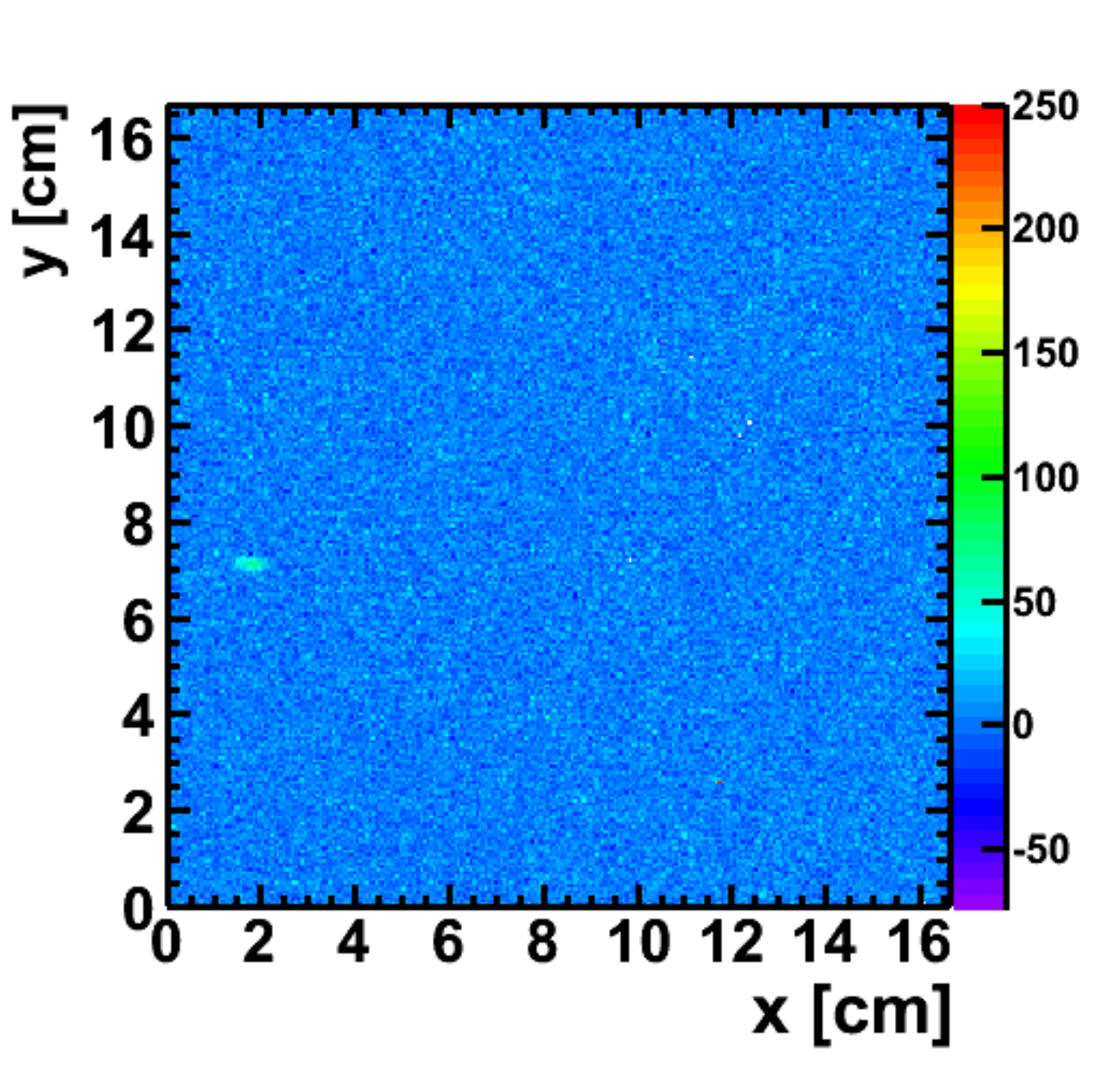}
\includegraphics[scale=0.20]{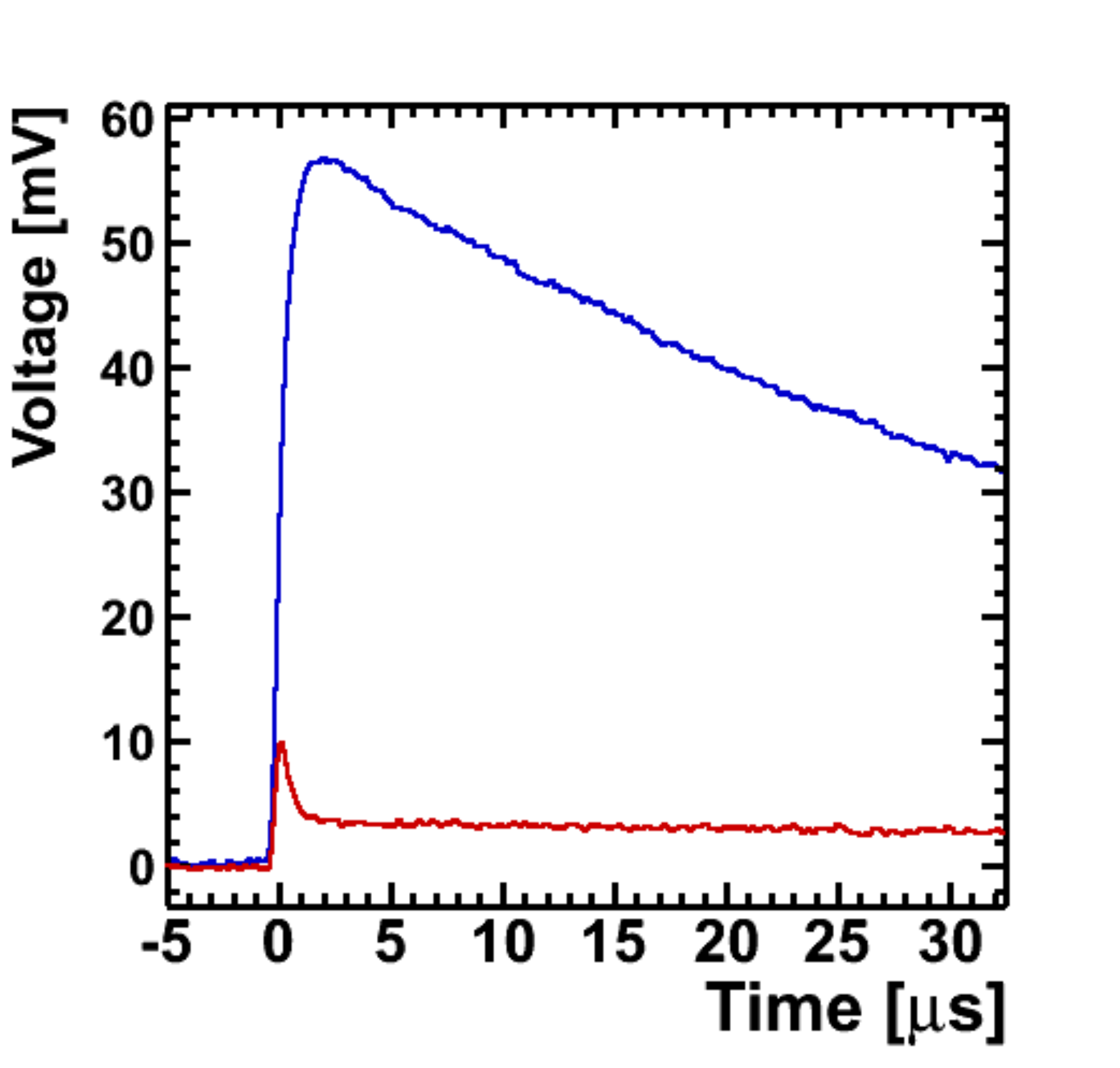}
\includegraphics[scale=0.20]{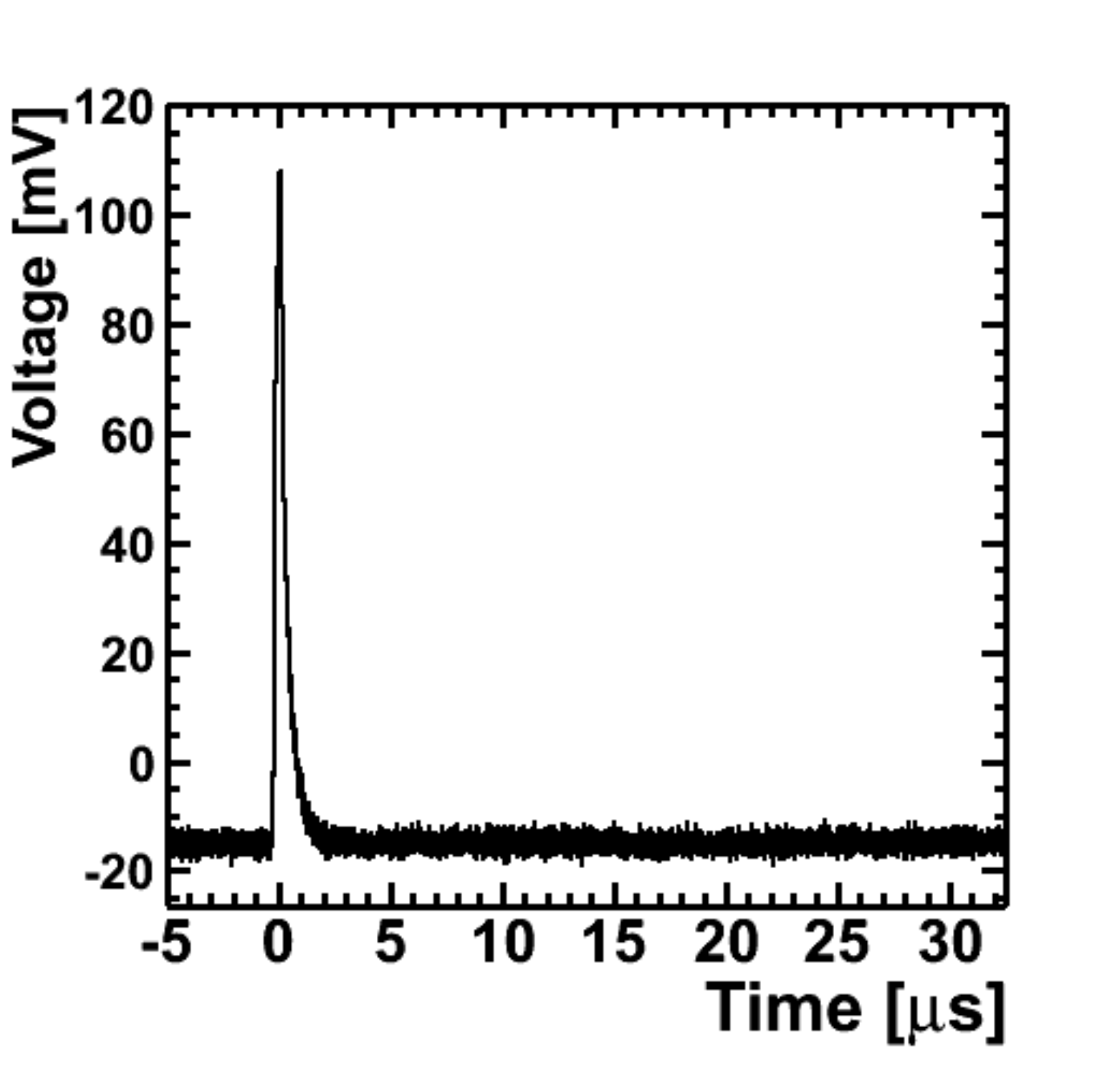}
\caption{
	Example figures of neutron recoils from $^{252}$Cf source showing
	graphical images of CCD and charge read-out system.
	Left: CCD image of $^4$He recoil. 
	Middle: Waveform from the Veto and Anode charge read-out. 
	Right: Waveform from the Ground-Mesh charge read-out.
	Top row is 100\% CF$_4$ and bottom is 12.5\% CF$_4$. \label{exampleNeutron}
}	
\end{center}
\end{figure}

The neutron source generates a large number of nuclear recoils, with $\mathcal{O}(1)$ 
neutron-induced nuclear recoils detected per 1~s exposure.  CCD images and charge waveforms of neutron-induced recoils in two events with
different gas mixtures are shown in Fig.~\ref{exampleNeutron}.  
A large number of electronic recoils 
from the high $\gamma$ flux are observed in the charge signals. However the CCD remains blind to these events due to their low primary ionization density. 
The source creates a much higher number of CCD-induced artifacts, 
likely from electronic recoils inside the active volume of the CCD
chip. Even with these large source activity issues, the cuts are sufficient for this analysis.

\begin{figure}
\begin{center}
\includegraphics[scale=0.30]{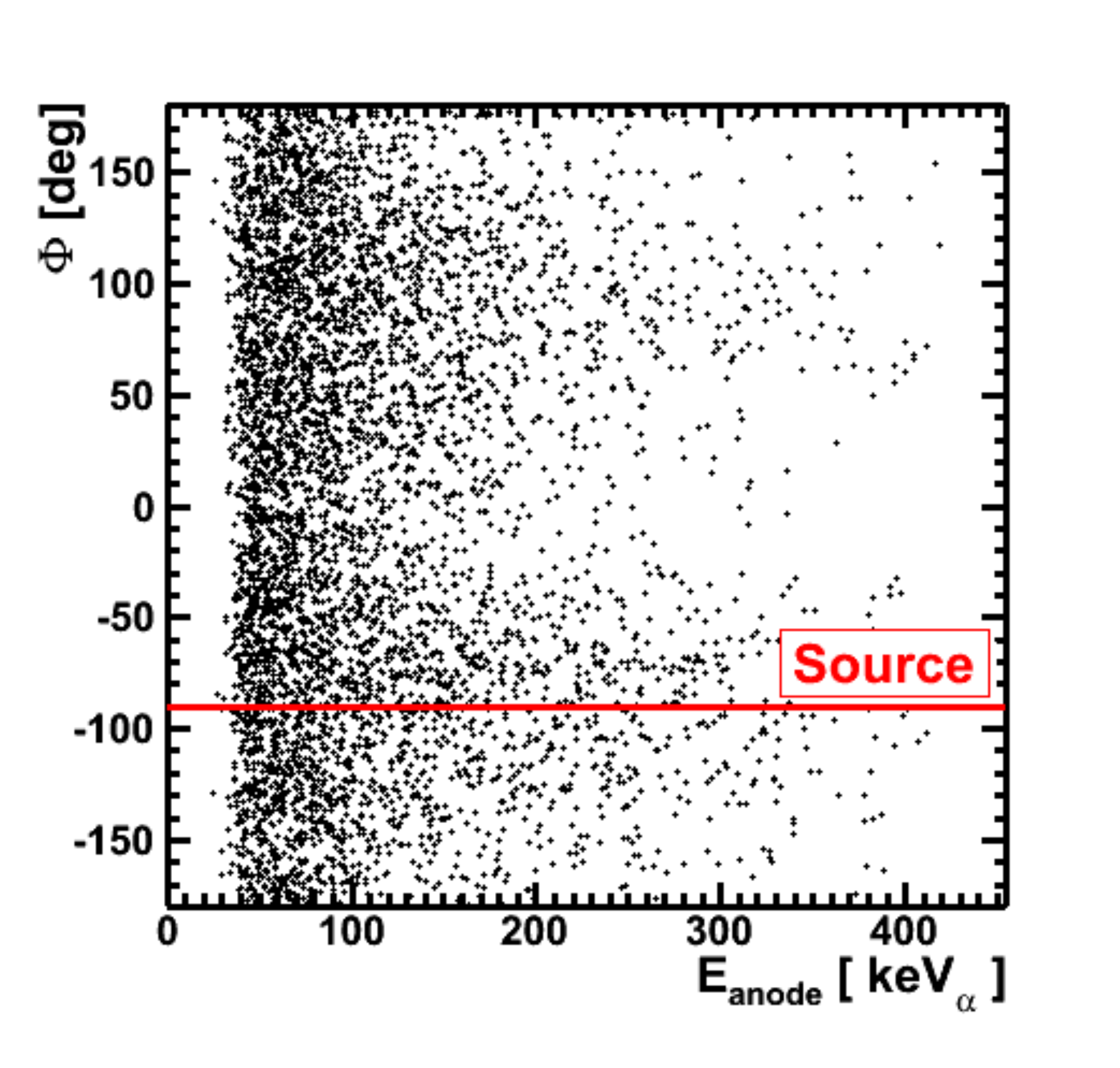}
\includegraphics[scale=0.30]{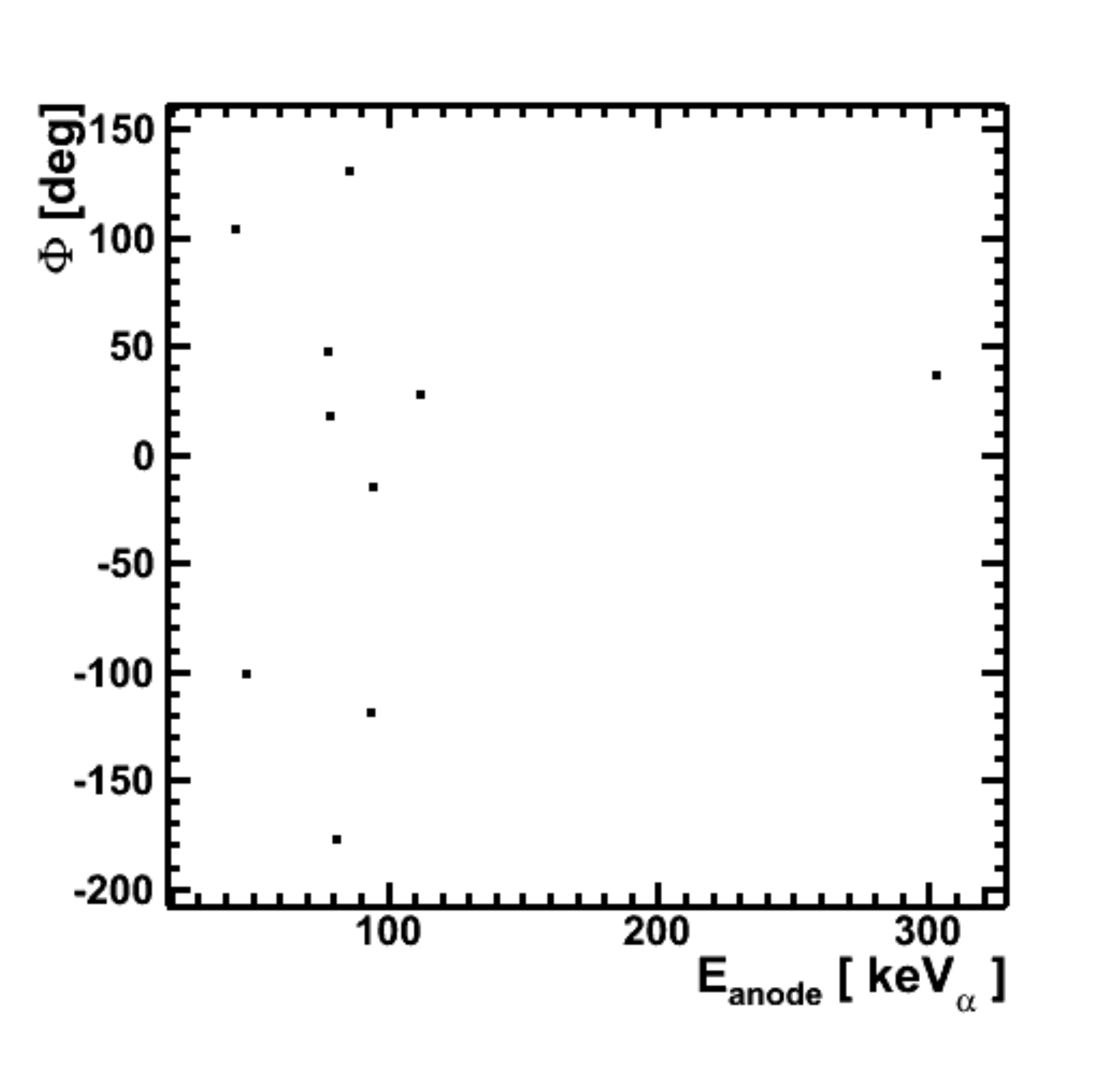}
\\
\includegraphics[scale=0.30]{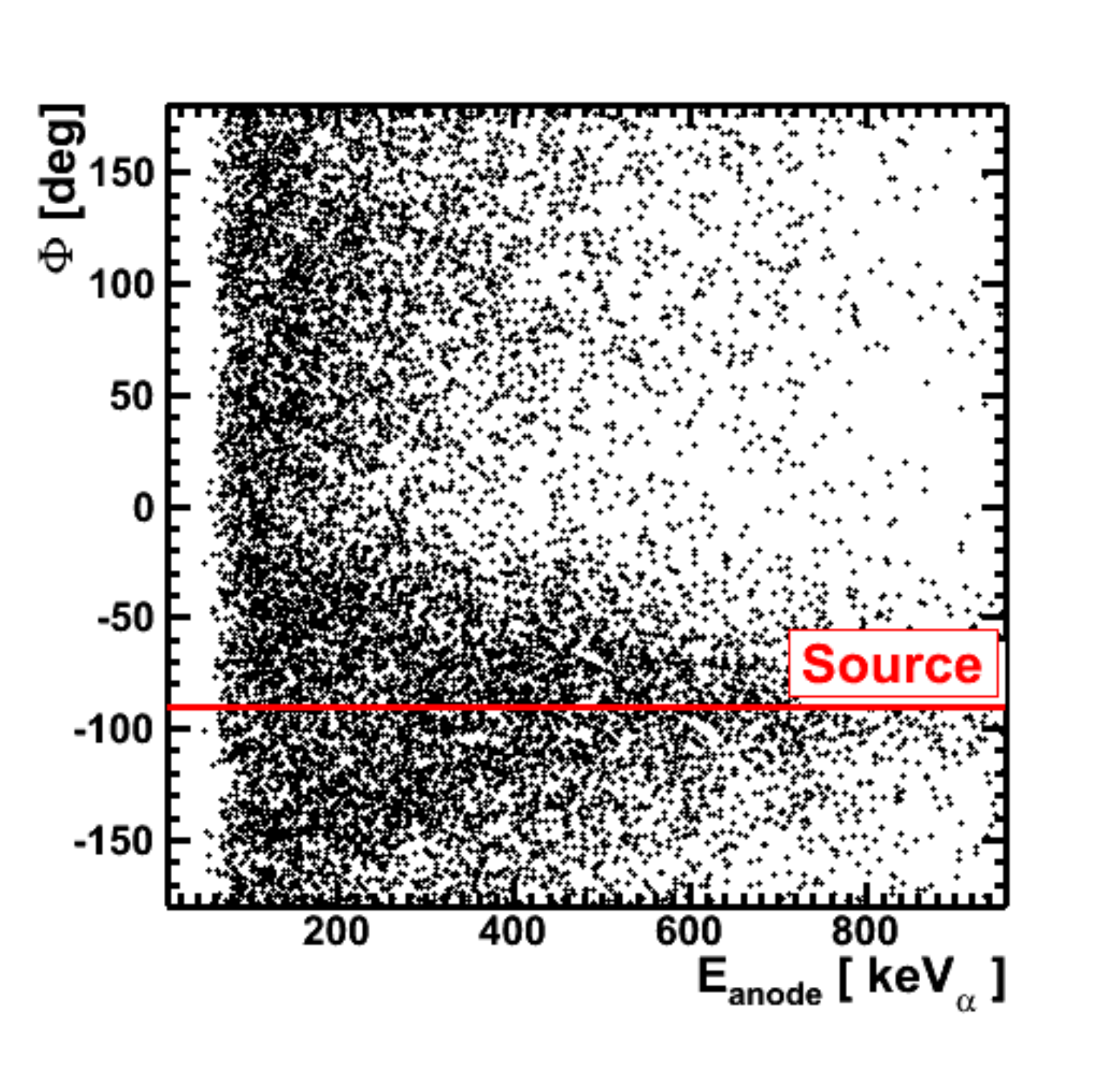}
\includegraphics[scale=0.30]{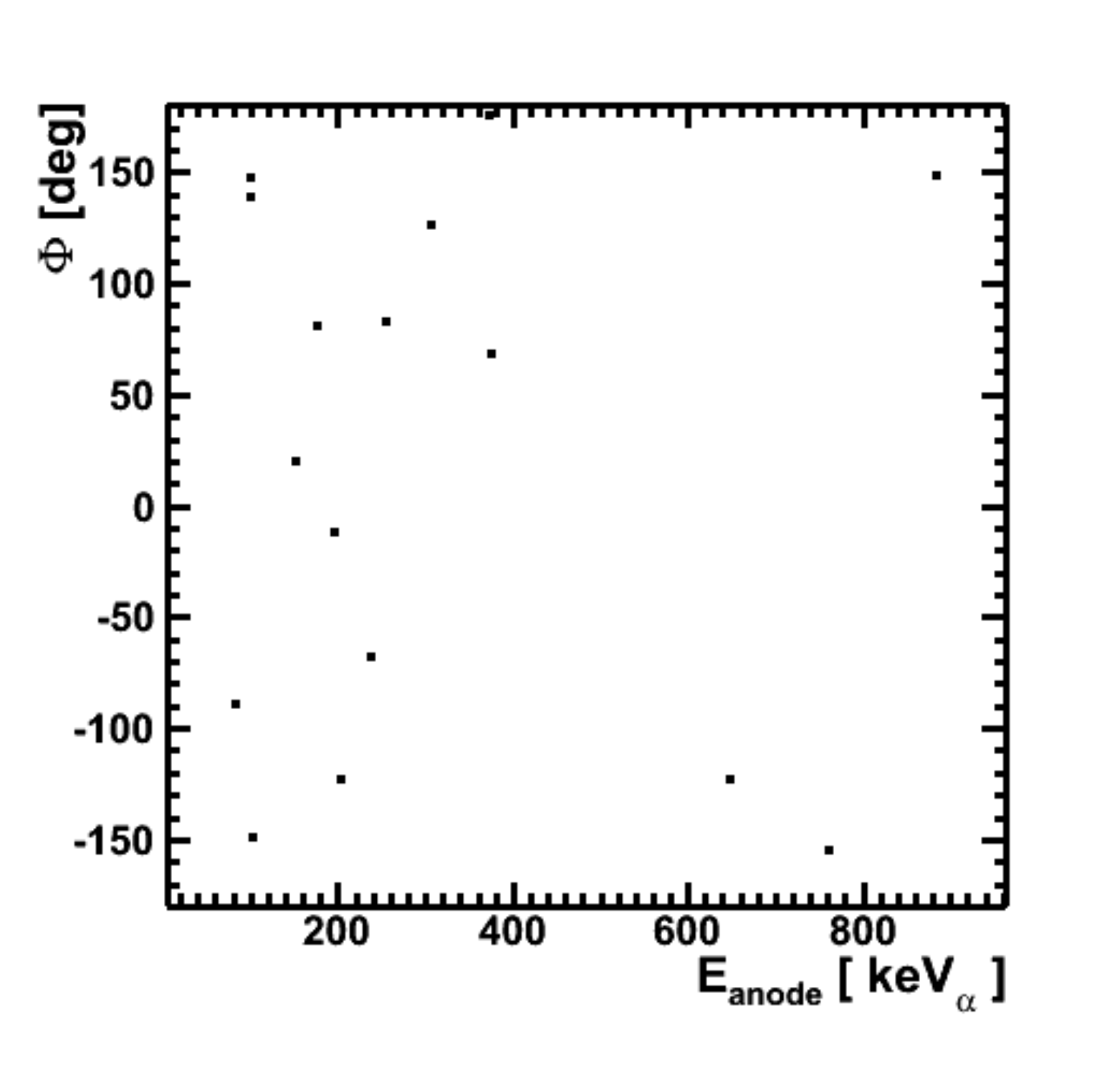}
\caption{
	Energy-Angular distribution for calibration and background
        runs for two gas mixtures. 
	Left: Calibration run with $^{252}$Cf source. Right: Background run.
	Top: 100\% CF$_4$ gas. Bottom: 12.5\% CF$_4$.
	The direction of the neutrons from the source is indicated by
        a red solid line at $-90$ degrees.
	The population at +90 degrees is due to incorrect reconstruction of the track direction. \label{angle}
}
\end{center}
\end{figure}

\begin{figure}
\begin{center}
\includegraphics[scale=0.325]{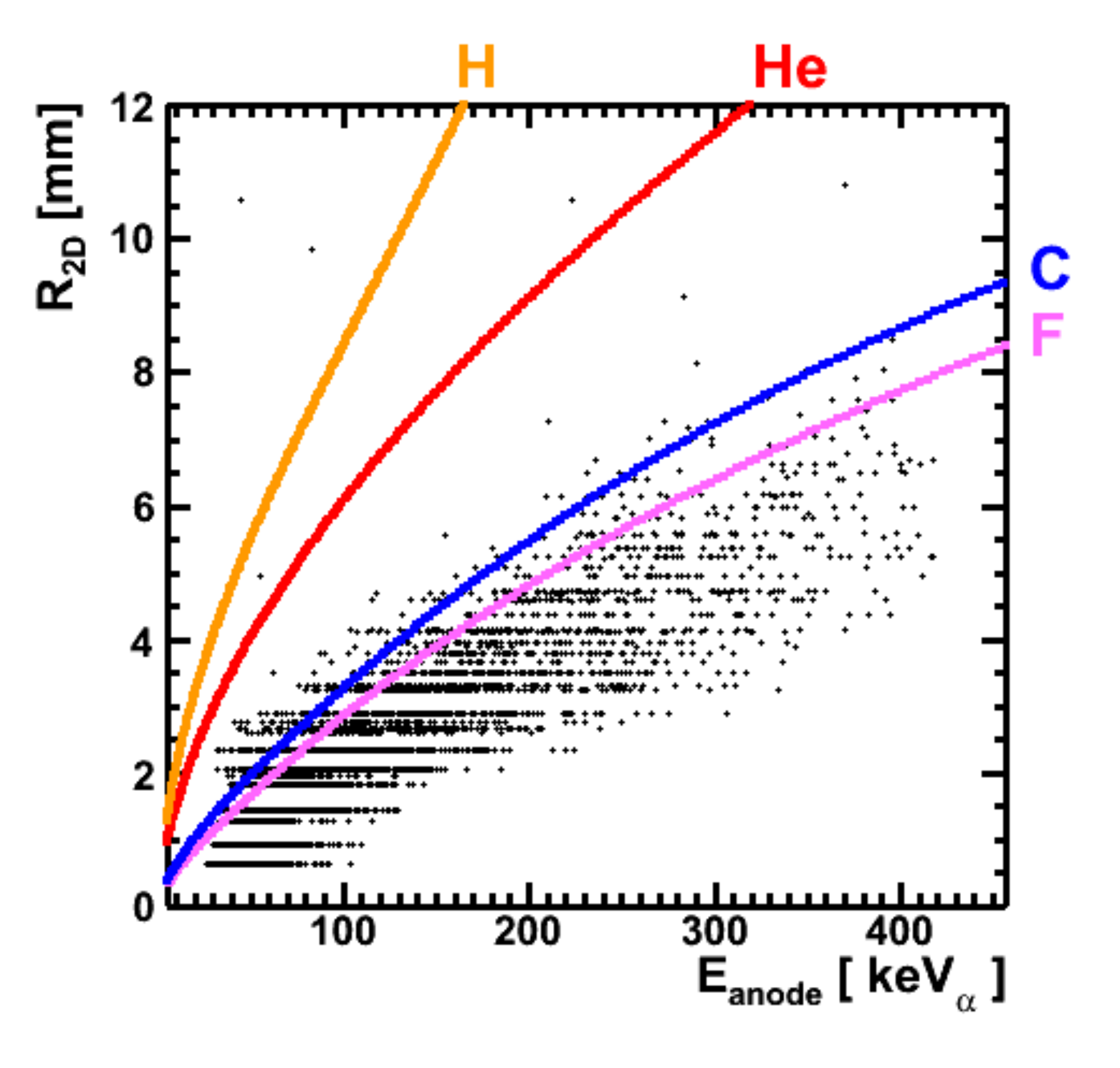}
\includegraphics[scale=0.325]{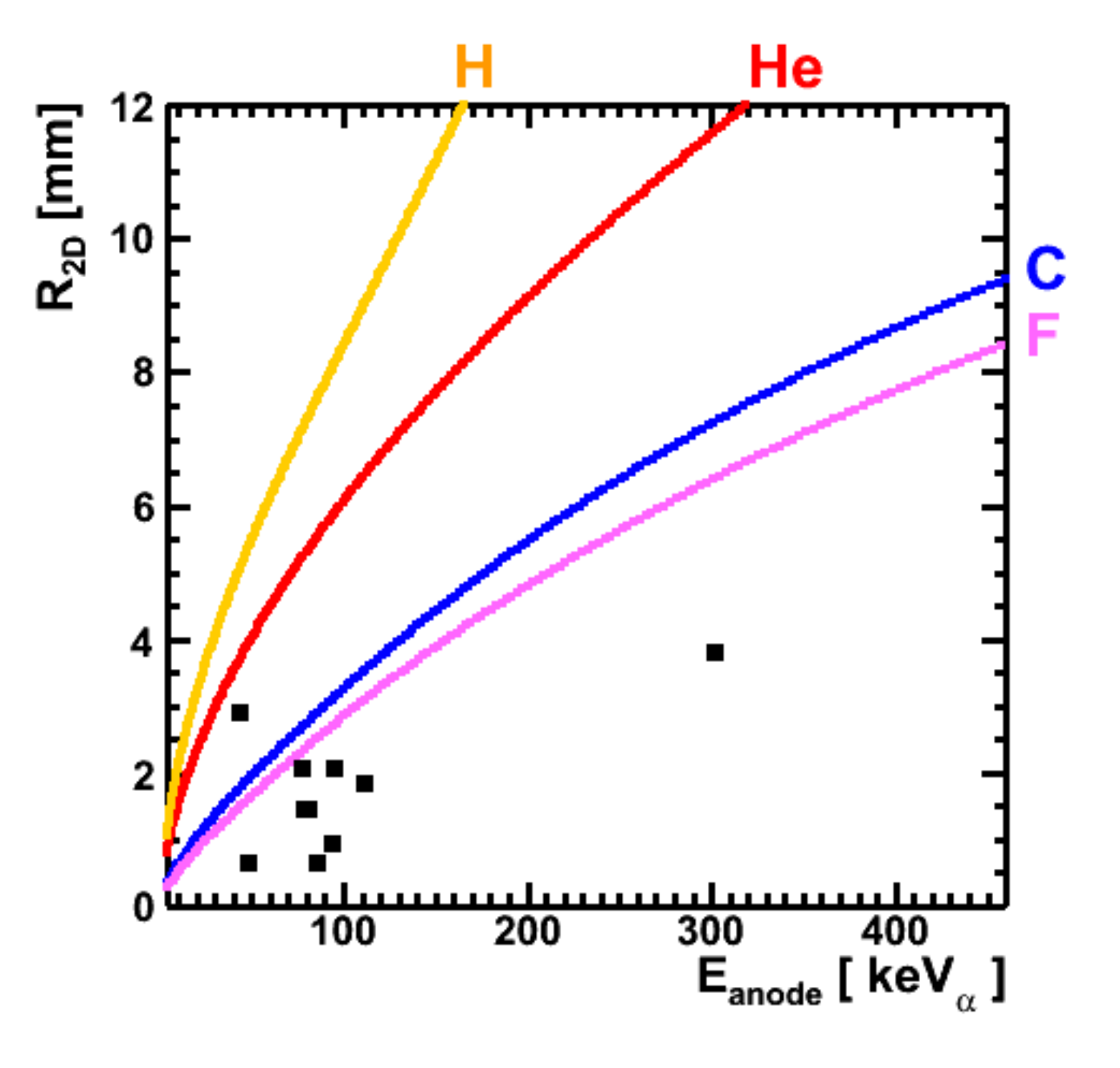}
\\
\includegraphics[scale=0.325]{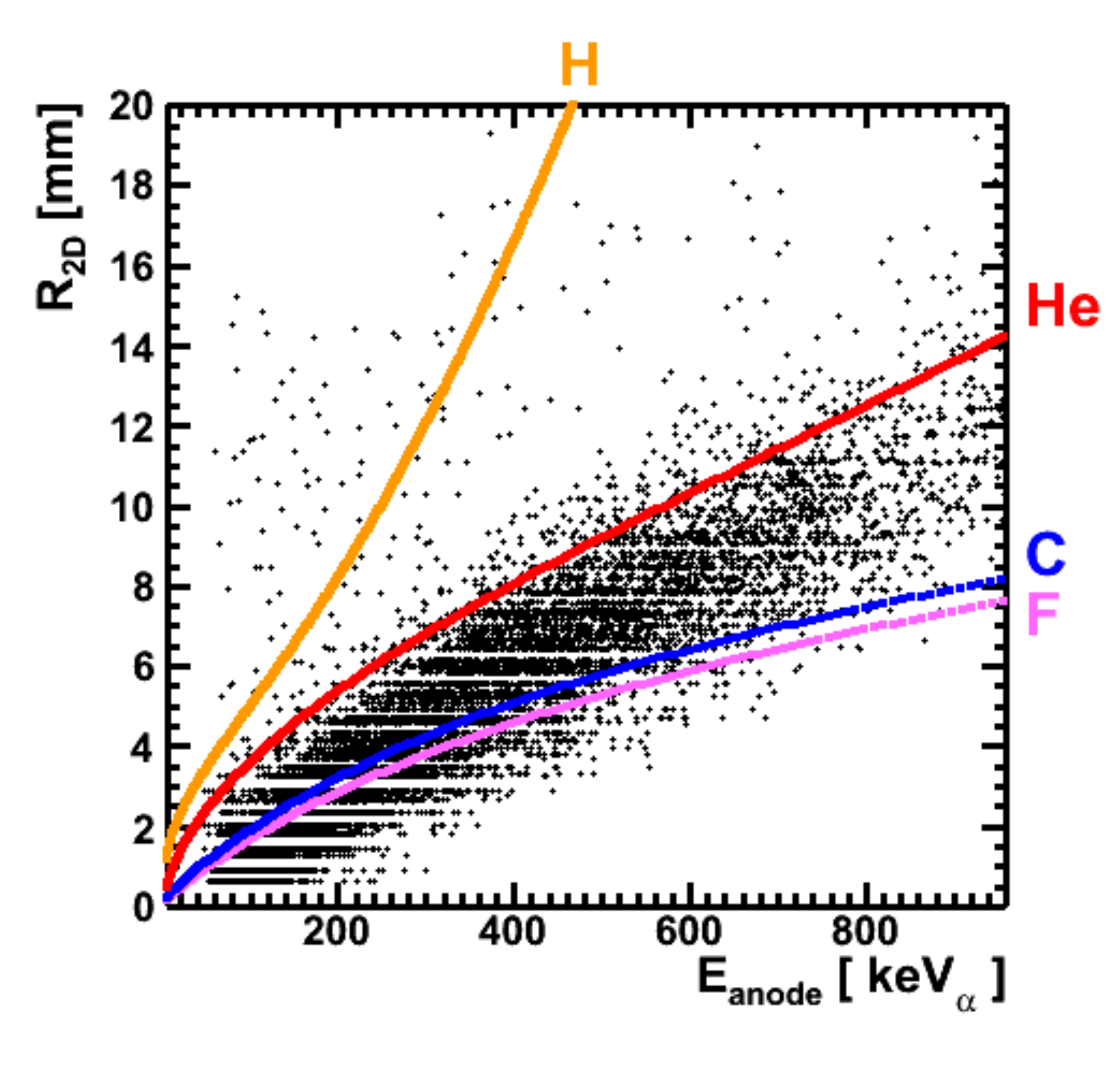}
\includegraphics[scale=0.325]{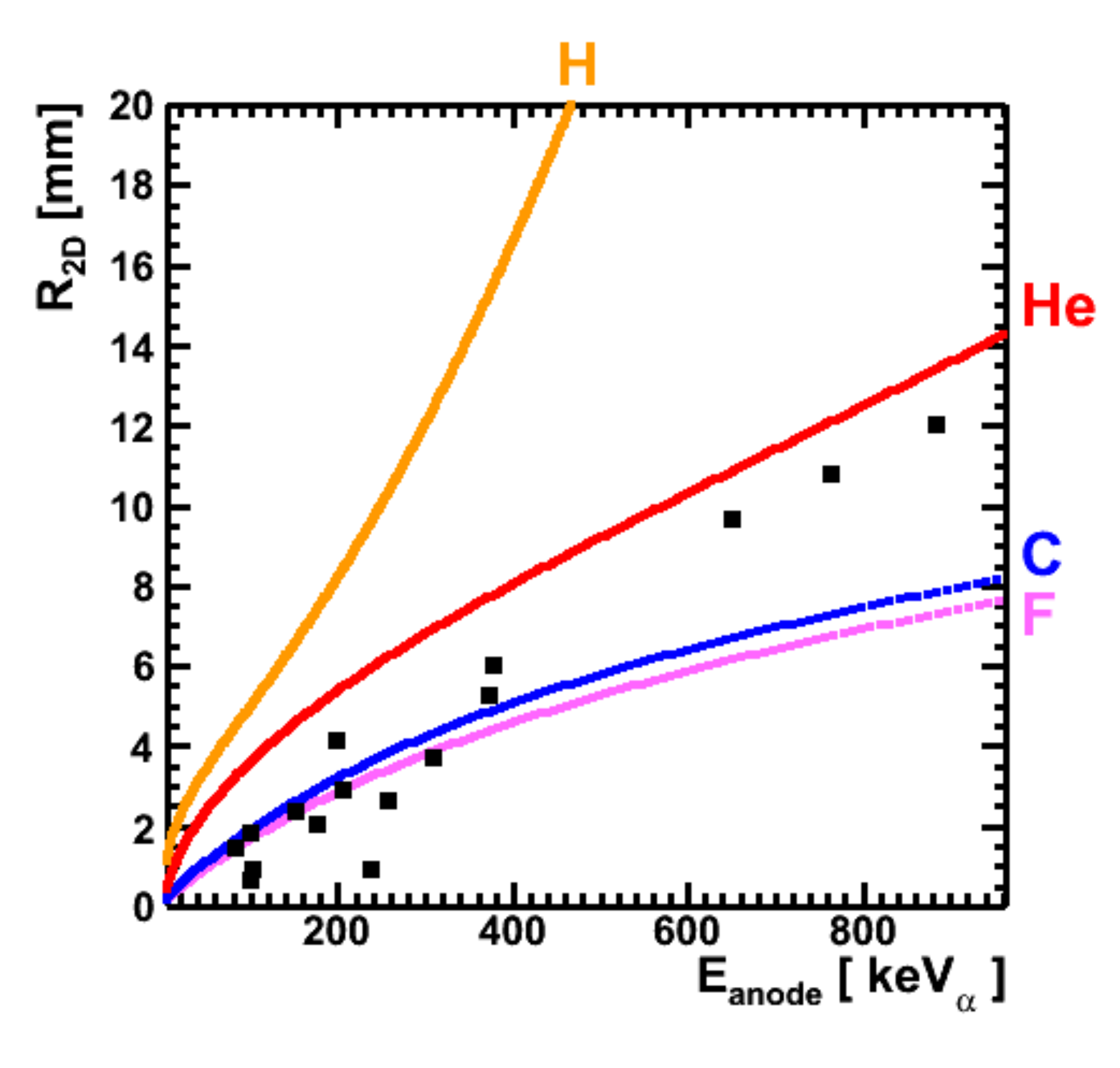}
\caption{
	Energy-Range distribution, calculated in two dimensions, for calibration and background runs with two different gas mixtures.
	Left: Calibration run with $^{252}$Cf source. Right: Background run.
	Top: 100\% CF$_4$ gas. Bottom: 12.5\% CF$_4$ mixture.
	Three-dimensional analytical prediction of the dE/dx curve for different nuclear elements of the gas mixture is shown in colored lines. \label{range}
}
\end{center}
\end{figure}

After applying all cuts, we use the measured track angles and ranges, Fig.~\ref{angle} and Fig.~\ref{range}, to evaluate the performance
of the detector and the analysis. The measured two-dimensional track range
generally falls just below the expected mean three-dimensional range
calculated with the publicly available program SRIM \cite{SRIM}.  For
the 12.5\% CF$_4$ mixture, the measured two-dimensional track range falls close to the expected
range for helium tracks.  This follows from the fact that helium recoils are kinematically favored across
most of the energy range measured here.  At lower energies, where both helium and fluorine recoils
are expected, the track ranges appear to be consistent with either
nucleus.  

We use the track angles to reconstruct the location of the neutron
source.  Tracks along the direction 
of the incoming neutrons should have a decreasing light profile, and
this ``head-tail effect'' can be used to reconstruct the sense of track
direction \cite{Dujmic:2007bd}. 
Two peaks are seen in the angular distribution of Fig.~\ref{angle}.  The larger peak at $-90$ degrees is along the 
true mean direction of the nuclear recoils. The second peak at $+90$ degrees is primarily from
events where the track direction was mis-reconstructed.  The large width of the distribution
is expected from elastic scattering kinematics.  Additionally, some nuclear recoils are from neutrons that have scattered
several times within the laboratory, and no longer appear to originate at the source position.

\begin{figure}
\includegraphics[width=7cm]{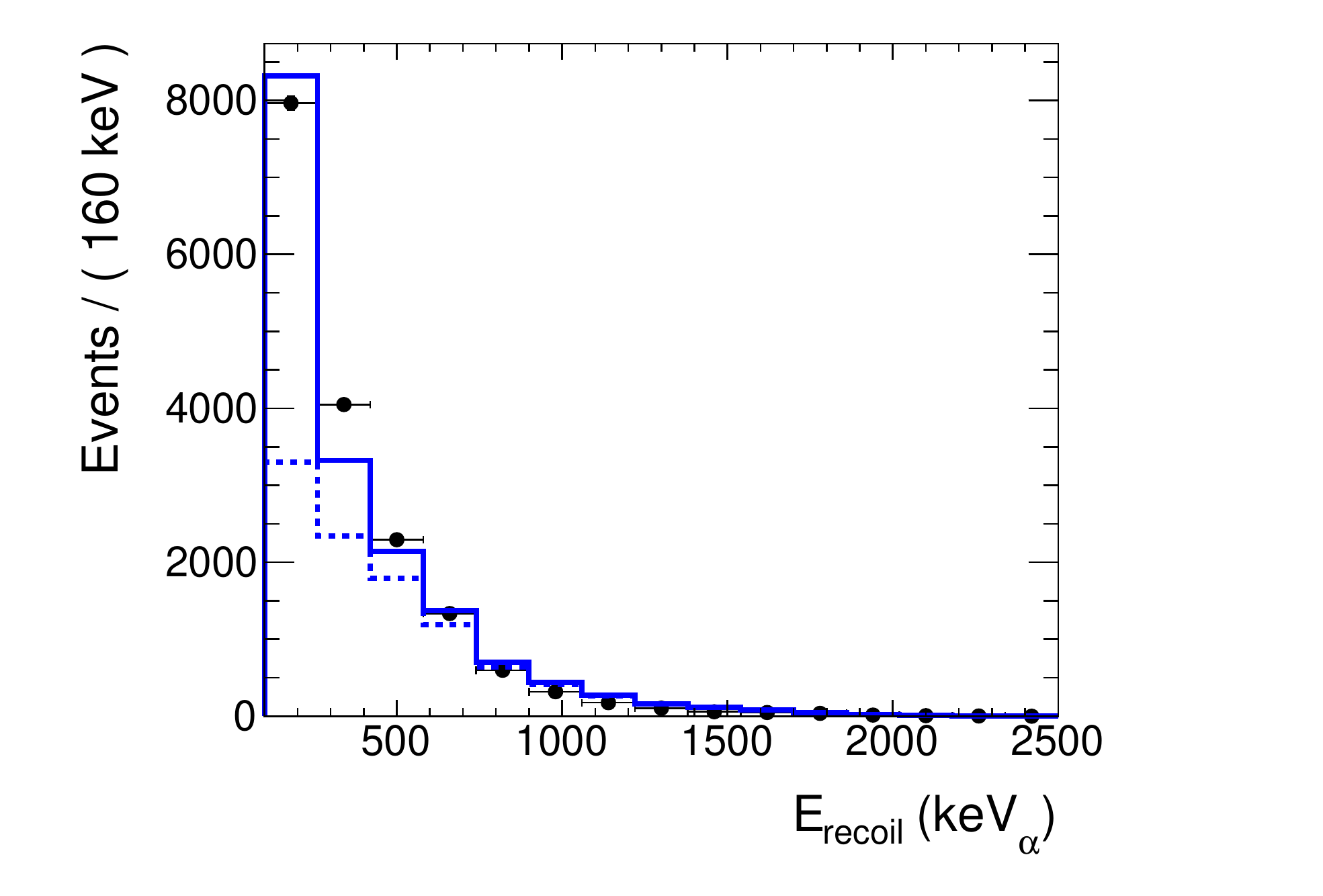}
\includegraphics[width=7cm]{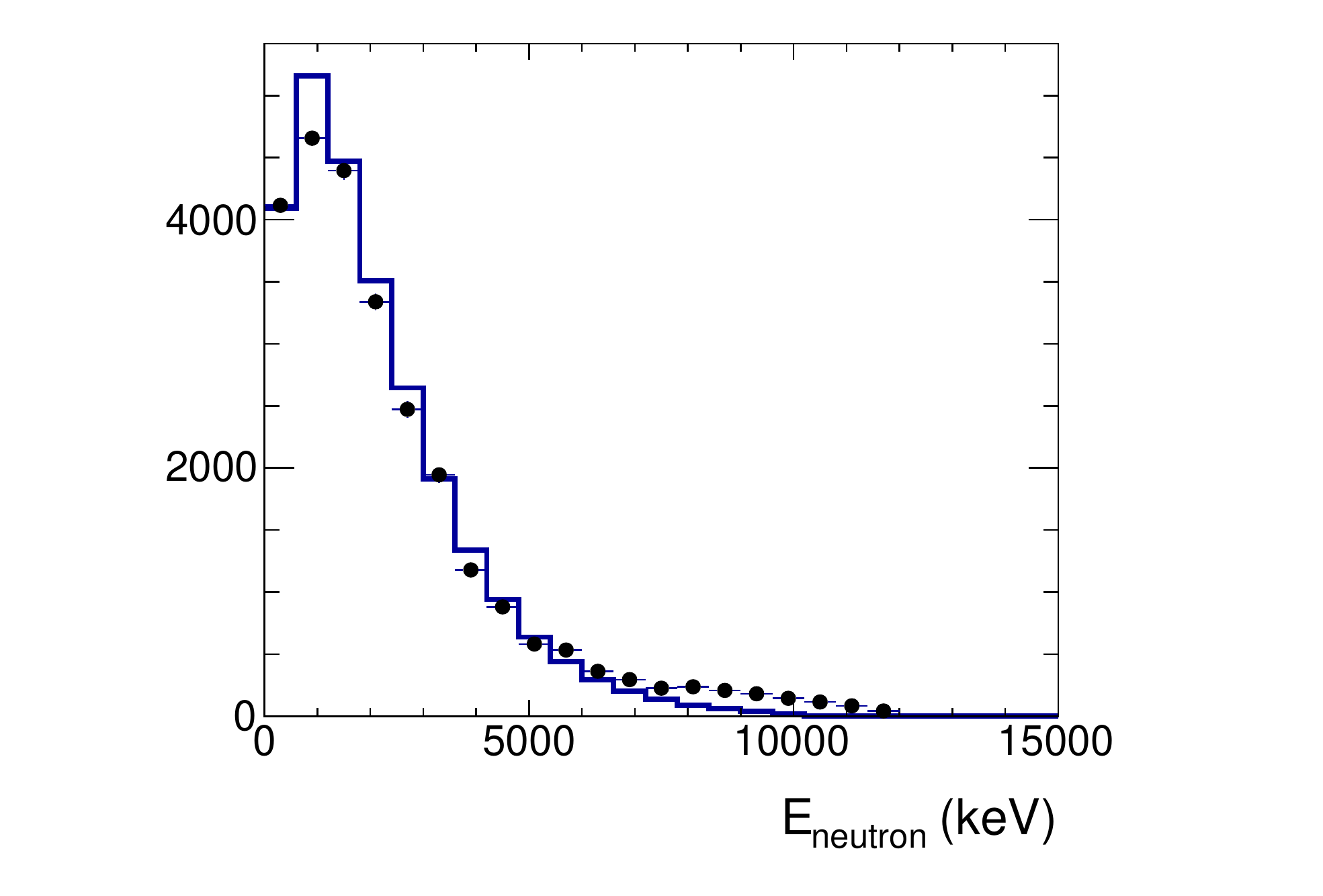}
\caption{ Left: A fit to a recoil spectrum (markers)  using a likelihood model based on helium and fluorine recoils.
The helium contribution (dashed) and total recoil distribution (full line) are shown as histograms.
Right: Unfolded neutron spectrum from a calibration run with $^{252}$Cf source (markers), and
a spectrum based on the $^{252}$Cf table~\cite{endf} (histogram). \label{fg::recoil_neutron_spectra}
}\end{figure}

A model of helium and fluorine recoils is fitted to the measured recoil spectrum in 
Fig.~\ref{fg::recoil_neutron_spectra}.  The fit uses the likelihood approach with the probability density functions for helium and fluorine 
based on simulation, and a $^{252}$Cf spectrum based on ENDF tables~\cite{endf}. 
We find  the fraction of helium recoil events to be $61\pm 1$~\%.  Another way to present this result is in terms of the unfolded neutron spectrum. We use the recoil data and the unfolding matrix, based on the simulation, to obtain the neutron spectrum as shown in 
Fig.~\ref{fg::recoil_neutron_spectra}. There is a reasonable agreement with the neutron spectrum from a $^{252}$Cf source that is based
on ENDF tables~\cite{endf}. In order to improve the agreement between the data and the simulation of the recoil and neutron spectra, 
we need to improve the simulation to include the efficiency loss due to sparks.  This requires
additional measurements, and a full simulation of the detector material which modifies the 
energy spectrum of neutrons entering the sensitive region.

\section{Surface Run}

With the DCTPC prototype, we have also taken data with no sources to measure cosmogenic, environmental
and detector backgrounds in a surface lab on the MIT campus.  The lab is on the ground 
floor of a two-story building. The lab has approximately 1~m concrete walls on all sides. After removing spark events, we obtain
112713 exposures in pure CF$_4$ and 103822 exposures in the 12.5\% CF$_4$
mixture.  As in the other runs, each exposure is 1~s, so this corresponds to 1.3 and 1.2 days of exposure, respectively.

In the pure CF$_4$ run, we identified 128 possible tracks in the CCD analysis.
Most of these tracks appear to be from CCD backgrounds such as noise artifacts
or ionization in the CCD. Of these 128 tracks, only 10 were matched to a charge trigger.
Several additional tracks had too much energy to be accurately measured by the charge
channels, although the pulses were collected.  These tracks are mostly low energy ($E<100 $ keV$_\alpha$)
and the ranges generally appear to be consistent with the expected values from SRIM for fluorine
recoils.  A single event has a range more consistent with helium or hydrogen tracks.

In the 12.5\% mixture run, we identified 71 potential tracks in the CCD analysis, with 16 having a matching
charge signal.   As with the pure CF$_4$ run, most of the potential tracks seen in the CCD are likely 
noise artifacts or tracks left directly in the silicon of the CCD chip.  As expected, we see more tracks
in this run due to the presence of helium.  The tracks seen here are also generally at higher energies than
were seen in pure CF$_4$.  This is to be expected if most of the tracks come from elastic scattering from
neutrons due to the favorable kinematics of neutron-helium scattering compared to neutron-fluorine scattering.
Fewer events at lower energies are seen due to the differing energy thresholds of the two gas mixtures.
There does not appear to be a favored direction of these tracks, although much higher statistics would
be necessary to confirm this.  The events with $E>300$ keV$_\alpha$ here have ranges consistent with helium recoils
but not carbon or fluorine, while most of the other tracks have ranges consistent with either carbon or fluorine 
nuclei or near vertical helium nuclei.

\section{Future Underground Measurement}

The first generation detector is being installed in the far-hall of
Double Chooz at 300 m.w.e. We can use the 13 neutrons per day measured
in the surface run to make an estimate for our event rate in the
far-hall. For the surface run, we measure a muon rate of 100~muons per
m$^{2}$ s$^{1}$. We assume the mean muon energy is $\sim$3 GeV. The
muon rate in the far-hall is 0.4~muons per m$^{2}$ s$^{1}$ with a mean
muon energy of 60~GeV\cite{DC}.  Using the power law scaling from
Ref. \cite{KamLAND}, we estimate $\sim$0.5 events per day.

\begin{figure}
\includegraphics[width=8cm]{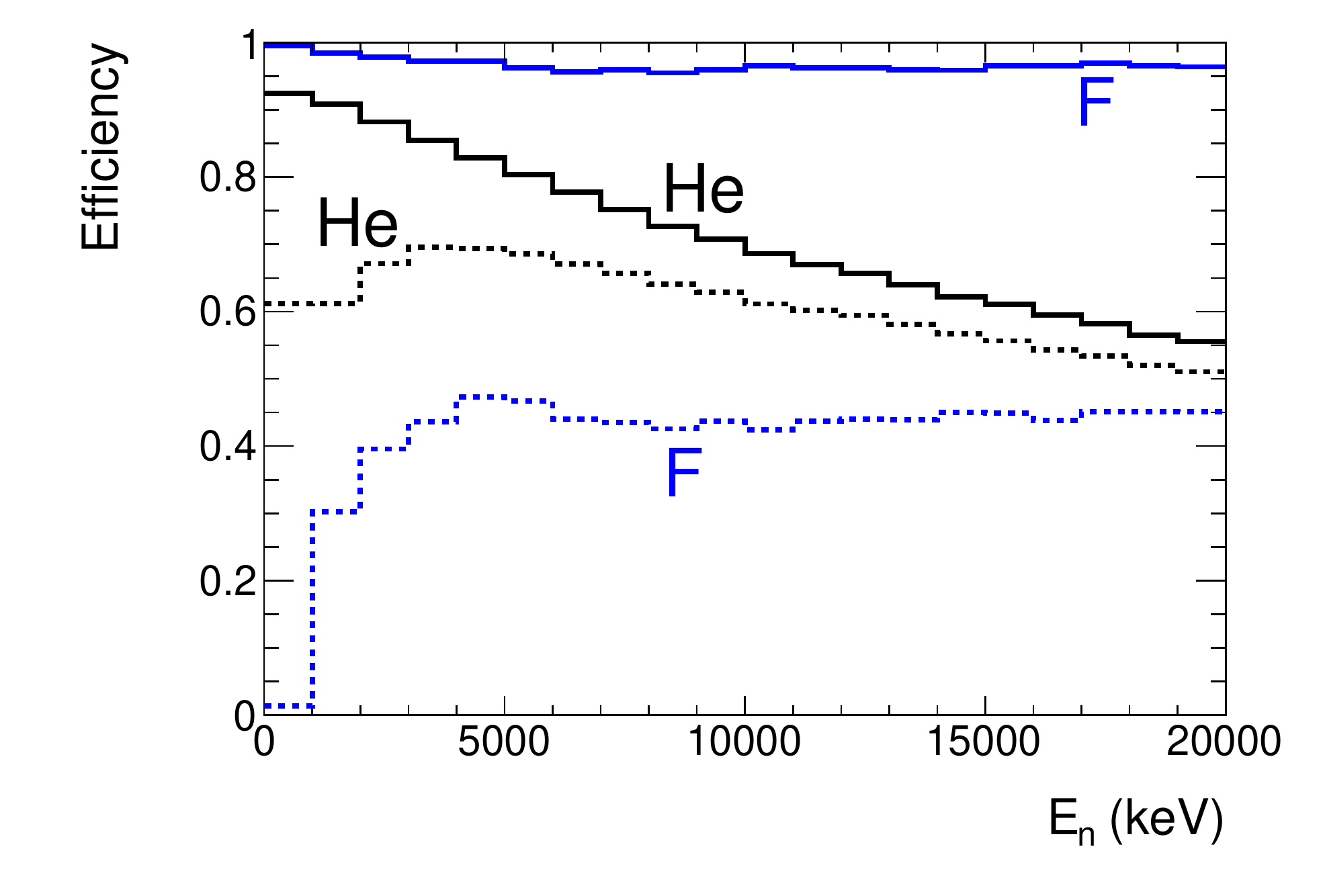}
\caption{ The efficiency for reconstructing helium and fluorine recoils as a function of neutron energy. The geometric efficiency (full line)
requires full track containment in the CCD view-field. The full efficiency (dashed line) assumes a step-function for the recoil 
reconstruction efficiency with thresholds given in the text.
\label{fg::eff_vs_neutron_energy} }
\end{figure}

A study of the detector efficiency, shown in
Fig.~\ref{fg::eff_vs_neutron_energy}, indicates that events collected in
this run is expected to sensitive to to neutrons
from 
($\alpha$, n) and muon interactions.
Fluorine and helium recoils are efficiently reconstructed above 
150 and 100~keV of recoil energy, respectively.  Since fluorine
recoils are shorter, they have a high probability to fit within CCD view-field. 
Geometric acceptance efficiencies are shown as full lines in Fig.~\ref{fg::eff_vs_neutron_energy}.
Helium recoils receive more energy from a collision with neutrons and leave longer tracks -- a combination that is favored by reconstruction algorithms.  Helium recoils
are more efficient at energies $\lesssim 20$~MeV. Fluorine appears more efficient if the detector is being
used for neutrons with the energy of tens of MeV's due to more compact
tracks.

\subsection{Conclusion}

We have modified the DMTPC dark matter detector design to produce a
2.8~liter, first-generation directional neutron detector.  
The primary change to the DMTPC design was to add helium as a target  for
the neutrons. We have demonstrated that a gas mixture of 75~Torr of CF$_4$ and
525~Torr of helium is capable of extracting the energy and angular distributions of the neutrons
from a $^{252}$Cf source.  
 
The sensitivity to higher energy neutrons and an estimate of 0.5
events per day at the Double Chooz far site indicates that the first
generation detector could make an interesting measurement at the far site on its own. The success of this detector motivates the construction of a full-sized DCTPC system.
This will consist of two 60~l detectors, located at the near and far
halls of Double Chooz, which can be interchanged for comparison of
systematics.  The goal is to measure the neutron flux, with
information on both energy and direction, as a function of time, in
order to tune the fast neutron simulation.

\section*{Acknowledgments}

The authors thank Prof. Peter Fisher, of MIT, and
the Double Chooz collaboration for valuable input.  JMC, KT and LW
are supported by the National Science
Foundation. DD is supported by the Department of Energy.
JL is supported by a Massachusetts Institute of
Technology Lyons Fellowship and the Institute for Soldier Nanotechnology.

\bibliographystyle{unsrt}

\bibliography{DCTPC_bib}

\end{document}